\documentclass[twocolumn]{aastex62}
\listofchanges

\usepackage{empheq,cases}

\received{}
\revised{}
\accepted{}
\submitjournal{AAS journals}

\shorttitle{New growth mechanism of dust grains in PPDs with MDWs}
\shortauthors{Taki et al.}
\begin{document}

\title{New growth mechanism of dust grains in protoplanetary disks with magnetically driven disk winds}

\correspondingauthor{Tetsuo Taki}
\email{taki@ea.c.u-tokyo.ac.jp}

\author[0000-0002-6602-7113]{Tetsuo Taki}
\affiliation{Center for Computational Astrophysics, National Astronomical Observatory of Japan, Osawa, Mitaka, Tokyo 181-8588, Japan}
\affiliation{School of Arts \& Sciences, University of Tokyo, 3-8-1, Komaba, Meguro, 153-8902 Tokyo, Japan}

\author{Koh Kuwabara}
\affiliation{Department of astronomy, University of Tokyo, 3-8-1, Komaba, Megro, 153-8902 Tokyo, Japan}

\author[0000-0001-8808-2132]{Hiroshi Kobayashi}
\affiliation{Department of Physics, Nagoya University, Nagoya, Aichi 464-8602, Japan}

\author[0000-0001-9734-9601]{Takeru K. Suzuki}
\affiliation{School of Arts \& Sciences, University of Tokyo, 3-8-1, Komaba, Meguro, 153-8902 Tokyo, Japan}
\affiliation{Department of astronomy, University of Tokyo, 3-8-1, Komaba, Megro, 153-8902 Tokyo, Japan}


\begin{abstract}
We discovered a new growth mode of dust grains to km-sized bodies in protoplanetary disks that evolve by viscous accretion and magnetically driven disk winds (MDWs).
We solved an approximate coagulation equation of dust grains with time-evolving disks that consist of both gas and solid components by a \replaced{one dimensional}{one-dimensional} model.
With the \replaced{collisional growth of dust grains}{grain growth}, all solid particles initially drift inward toward the central star by the gas drag force.
However, the radial profile of gas pressure, $P$, is modified by the MDW that disperses the gas in an inside-out manner.
\replaced{As a result}{Consequently}, a local concentration of solid particles is created by the converging radial flux of drifting dust grains at the location with the convex upward profile of $P$.
When the dimensionless stopping time, \deleted{or the Stokes number, }St, there exceeds unity, the solid particles spontaneously reach the growth dominated state because of the positive feedback between the suppressed radial drift and the enhanced accumulation of dust particles that drift from the outer part.
\deleted{These consecutive processes proceed even for spherical dust grains under the Epstein gas drag.}
Once the solid particles are in the drift limited state, the \replaced{above mentioned}{above-mentioned} condition of St $\gtrsim 1$ for the dust growth is equivalent with
\begin{equation}
  \Sigma_{\rm d}/\Sigma_{\rm g}\gtrsim \eta, \nonumber
\end{equation}
where $\Sigma_{\rm d}/\Sigma_{\rm g}$ is the dust-to-gas surface-density ratio and $\eta$ is dimensionless radial pressure-gradient force.
As a consequence of the successful growth of dust grains, a ring-like structure containing planetesimal-sized bodies is formed at the inner part of the protoplanetary disks.
Such a ring-shaped concentration of planetesimals is expected to play a vital role in the subsequent planet formation.
\end{abstract}



\section{Introduction}
\label{sec:introduction}
Recent observations by large astronomy facilities such as the Atacama Large Millimeter/Submillimeter Array (ALMA) and Subaru telescope revealed varieties of complicated structures of protoplanetary disks \citep[PPDs hereafter; ][]{Hashimoto2011a, Casassus2013a, Fukagawa2013a, ALMA-Partnership2015a, Benisty2015a, Andrews2011a, Andrews2018b, Andrews2018a, Akiyama2019a}.
It is important to understand how planets are formed in evolving PPDs with various complicated structures.

In addition to viscous accretion \citep[e.g.,][]{1974MNRAS.168..603L}, there are various processes that affect the evolution of PPDs, such as photoevaporation \citep{Shu1993a, Owen2012a, Ercolano2015a, Hollenbach2017a}, the magnetically driven disk wind (MDW) \citep{2006A&A...453..785F,Suzuki2009a,2013ApJ...769...76B,2013A&A...550A..61L,Suzuki2016a}, and non-ideal MHD effects \citep{Sano2004a, Bai2013a, Simon2015a, Suriano2018a,Suriano2019}.
An interesting aspect of these processes is that they potentially create characteristic features in the surface density of the gas because different processes effectively operate at different locations.

In usual conditions, such as the minimum-mass solar nebula \citep[MMSN hereafter;][]{Hayashi1981a}, the gas pressure generally decreases with the radial distance from a central star.
In such circumstances, one of the severe obstacles against the planet formation is the infall of solid particles toward the central star by the gas drag.
The gas component rotates with a sub-Keplerian velocity because of the outward pressure gradient force to the radial force balance, while the solid component tends to rotate with the Keplerian velocity.
Therefore, solid particles feel headwind from the gas so that their rotation velocity is decelerated.
As a result, these solid particles move toward the central star \citep{1972fpp..conf..211W,Adachi1976a,Weidenschilling1977a}.
The timescale of this inward drift is significantly shorter than the growth time of solid particles when we assume the direct sticking of solid particles as the growth mechanism.
This obstacle is often called ``the radial drift barrier''.

There have been various mechanisms introduced to overcome the radial drift barrier.
The rapid formation of large-sized bodies is one possible solution to this barrier.
A possible path is two-fluid instability between the gas and solid components, which is called streaming instability \citep{Youdin2005a}.
The streaming instability forms dense clumps of solid particles quickly \citep{Johansen2007a}.
These dense clumps eventually collapse to objects with a size of $\sim 10^2$ km by the self-gravity \citep{Johansen2012a}.
These large objects are no longer affected strongly by the background gas flow, because they are too massive to be perturbed by the gas drag.
Considering the internal density evolution of solid particles is an alternative way to the rapid formation of large-sized bodies.
\citet{Okuzumi2012a} computed the evolution of dust mass and their internal density simultaneously, and they found that fluffy dust aggregates can quickly grow into large-sized bodies rather than falling toward the central star.

Substructures of gaseous disks are another possible solution to the radial drift barrier.
For example, a local maximum of gas pressure, which is often called a ``pressure bump'', is a  promising location to halt the radial drift of solid particles \citep{1972fpp..conf..211W,2003ApJ...583..996H,2003ApJ...598.1301H,Taki2016a}.
This is because the direction of the radial drift of dust particles is the same as the direction of the pressure gradient, which is derived from the radial force balance of the gas.
Several candidates that yield pressure bumps are proposed: some examples are the inner edge of PPDs \citep{2003ApJ...583..996H}, the inner edge of dead zones \citep{2008A&A...491L..41L,2010A&A...515A..70D,Suzuki2010a}, which are inactive with respect to magnetorotational instability \citep[MRI hereafter;][]{velikhov1959stability,1961hhs..book.....C,1991ApJ...376..214B}, and snowlines \citep{2007ApJ...664L..55K}.

The MDW potentially creates a pressure bump near the inner edge of PPDs \citep{Suzuki2010a, Takahashi2018a}.
The mass loss timescale owing to the MDW is scaled by the local Keplerian time.
Therefore, the MDW generally disperses the gas component of a PPD in an inside-out manner.
In other words, an inner cavity of the gaseous disk is expected to from by the MDWs.
\citet{Takahashi2018a} calculated the evolution of the surface density of both gas and solid components simultaneously with MDWs.
They confirmed that pressure bumps formed by MDWs halt the radial drift of dust particles, and showed that the dust surface density gives a ring-hole configuration.
An interesting aspect of the pressure bump formed by the MDW is that it moves outward with time.
The dust ring also moves outward as the location of the pressure maximum moves.

The radial drift velocity of solid particles is a function of their size and the solid-to-gas density ratio \citep{Nakagawa1986a}.
Namely, the growth of dust particles is also important in forming ring-hole structures, in addition to the accumulation of the solid component.
Since the accumulation of solid particles effectively occurs at a pressure bump, the growth timescale of solid particles is significantly shorter there than that in other parts of PPDs.

The main focus of the present paper is to investigate the size evolution of dust particles, which was not considered in \citet{Takahashi2018a}, in PPDs with MDWs.
We calculate a coagulation equation of solid particles in time-evolving PPDs with MDWs.
Although previous works have mainly focused on pressure bumps, we further pursue a new mechanism that piles up dust particles in PPDs.

The construction of this paper is as follows.
In Sect. \ref{model} we describe the equations and simulation settings.
In Sect. \ref{results} we show the simulation results of different cases and explain a newly discovered growth mechanism of solid particles.
In Sect. \ref{discussion} we discuss caveats of our model and implications for the observation of PPDs and the formation of planetary systems.
Our conclusions are presented in Sect. \ref{conclusions}.

\section{Model} \label{model}
We solve a coagulation equation of dust grains with the time-evolving
surface densities of the gas and solid components under the
axisymmetric approximation. We describe each equation below.

%%%%%%%%%%%%%%%%%%%%%%%%%%%%%%%%%%%%%%%%%%%%%%%%%%%%%%%%%%%%%%%%%%%%%%%
\subsection{Gas Surface Density}\label{sect:gas}
The time evolution of gas surface density, $\Sigma_{\rm g}$,  with MDWs can be written as
\begin{equation}
 \frac{\partial \Sigma_{\rm g}}{\partial t} +
 \frac{1}{r}\frac{\partial}{\partial r} \left(r \Sigma_{\rm g} v_{{\rm
     g},r}\right) +\left(\rho_{\rm g} v_{{\rm g},z}\right)_{\rm w} = 0,
 \label{eq:gas}
\end{equation}
where $r$ is the radial distance from a central star and $v_{{\rm g},r}$ and $v_{{\rm g},z}$ are the $r$ and $z$ (vertical) components of the velocity of gas.
The subscript ``w'' of the third term represents the mass loss by MDWs.
We neglect the backreaction from the solid component in eq.~(\ref{eq:gas}).
$\Sigma_{\rm g}$ is related to the gas density at the midplane,
$\rho_{\rm g}$, via
\begin{equation}\label{eq:rhomidplane}
\rho_{\rm g} = \frac{1}{\sqrt{2\pi}} \frac{\Sigma_{\rm g}}{h_{\rm g}}.
\end{equation} 
Here, the vertical scale height of gas,
\begin{equation}
h_{\rm g} = \frac{c_{\rm s}}{\Omega_{\rm k}},
\end{equation}
is derived from the sound speed, $c_{\rm s}$, and the Keplerian frequency, $\Omega_{\rm k} = \sqrt{G M_{\star}/r^3}$, where $G$ is the gravitational constant and $M_{\star}$ is the mass of the central star.
Throughout this paper, we consider a star with one solar mass,
$M_{\star} = 1M_{\odot}$.
We adopt the temperature structure that is determined by the radiative equilibrium, $ T= 280 \times \left(r/1 \  \rm au \right)^{-1/2} \rm{K} $ \citep{Hayashi1981a}.
From this we obtain $c_{\rm s} =  \sqrt{k_{\rm B} T / \mu m_{\rm u}} = 9.9 \times 10^{4} \left(r/1 \  \rm au \right)^{-1/4} {\rm cm\ s^{-1}}$, where $k_{\rm B}$ is the Boltzmann constant, $m_{\rm u}$ is the unified atomic mass unit, and $\mu = 2.34$ \citep{Hayashi1981a} is the mean molecular weight.
From these dependences, the scale height depends on $r$ as $h_{\rm g} \propto r^{5/4}$.

The second term of eq.~ (\ref{eq:gas}) denotes the radial flow of the gas and it is calculated from the conservation of the angular momentum in an annulus of a disk \citep{Suzuki2016a} as
\begin{equation}
 r \Sigma_{\rm g} v_{{\rm g},r} = -\frac{2}{r \Omega_{\rm k}}
 \left[\frac{\partial}{\partial r} \left(r^{2} \Sigma_{\rm g}
   \overline{\alpha_{r \phi}} c_{\rm s}^{2}\right) + r^{2}
   \overline{\alpha_{\phi z}} (\rho_{\rm g} c_{\rm s}^{2})_{\rm mid}
   \right],
\label{eq:vgr}
\end{equation}
where the subscript, ``mid'', stands for the midplane and $\overline{\alpha_{r,\phi}}$ and $\overline{\alpha_{\phi, z}}$ are dimensionless parameters after \citet{Shakura1973a}.
$\overline{\alpha_{r\phi}}$ is an effective turbulent viscosity, and $\overline{\alpha_{\phi z}}$ is a magnetic braking stress, namely the torque exerted from the MDW, where $\overline{\cdots}$ denotes that the density-weighted averages are taken over the $\phi$ and $z$ directions \citep[see][for the mathematical definitions]{Suzuki2016a}.
These two parameters, $\overline{\alpha_{\phi z}}$ and $\overline{\alpha_{r \phi}}$, respectively determine the outward transport and removal of the angular momentum from a disk that induce the accretion of gas to the central star.

We consider two types of the parameterization for the wind torque: (i) $\overline{\alpha_{\phi z}}$ is constant in the whole region; and (ii) $\overline{\alpha_{\phi z}}$ depends on the local gas surface density.
We name (i) constant torque and (ii) $\Sigma$-dependent torque from now on, following \citet{Suzuki2016a}.
The wind torque is expressed as
\begin{equation}
 \label{eq:alpha_pz}
 \overline{\alpha_{\phi z}} = \overline{\alpha_{\phi z, 0}} \left(\frac{\Sigma_{g}}{\Sigma_{g,0}}\right)^{-l}.
\end{equation}
We assume that $\overline{\alpha_{\phi z, 0}} = 1.0\times 10^{-4}$ and $l = 0$ for the constant torque case, and $\overline{\alpha_{\phi z, 0}} = 1.0 \times 10^{-5}$ and $l=0.66$ for the $\Sigma$-dependent case.
\citet{Bai2013a} reported that $\overline{\alpha_{\phi z}} \sim 10^{-5}...10^{-3}$ with a negative dependence on the plasma $\beta$, $\overline{\alpha_{\phi z}} \propto (8\pi(\rho c_s^2)_{\rm mid}/B_{z}^2)^{-0.66}$, by local MHD simulations.
The constant torque case corresponds to the case that the plasma $\beta$ is uniform for time and location \citep{Bai2016a}.
The $\Sigma$-dependent torque case corresponds to the case that the vertical magnetic flux is preserved at each location even though the gas surface density decreases.
We test these two extreme cases in this paper.
\added{
We note that the adopted value of $\overline{\alpha_{\phi z}}$ should be tested by global treatments because the local shearing box approximation cannot capture angular momentum in principle \citep{1995ApJ...440..742H}.
Recent 2D global simulations by \citet{2020ApJ...896..126G} reported a typical value of $\overline{\alpha_{\phi z}}\sim 10^{-4}$, whereas it shows large scatter.
Therefore, our adopted $\overline{\alpha_{\phi z}}$ is probably reasonable within an order-of-magnitude argument.
}

The third term of eq.~(\ref{eq:gas}) indicates the mass loss by the MDW, and the mass loss rate is adopted from the local MHD simulations in \citet{Suzuki2010a} as follows,
\begin{equation}
\label{Cw}
\left(\rho_{\rm g} v_{{\rm g},z}\right)_{\rm w} = C_{\rm w} \left(\rho_{\rm g} c_{\rm s} \right)_{\rm mid},
\end{equation}
where $C_{\rm w}$ is the dimensionless mass flux of the MDW.
We employ the typical value of $C_{\rm w} \sim 10^{-5}$.

% response to 2-1
\added{
We here discuss the relation of our parametrization to the different driving  mechanisms of the MDWs, although we do not specify them in detail in our model calculations.
$\overline{\alpha_{\phi z}}$ generally determines the relative contribution of the magnetic tension by poloidal fields in driving MDWs.
When $\overline{\alpha_{\phi z}}$ is large, the disk wind is accelerated primarily by magnetic tension force, which is called a magneto-centrifugal driven disk wind \citep{Blandford1982a}.
In the opposite extreme limit of $\overline{\alpha_{\phi z}}=0$, the disk wind is driven by the pressure of toroidal magnetic fields \citep{1997ApJ...474..362K} or magneto-turbulence \citep{Suzuki2009a}.
The MCW regime is studied by the constant $\overline{\alpha_{\phi z}}\ne0$ and $\Sigma$-dependent $\overline{\alpha_{\phi z}}$ prescriptions mentioned above, while the pressure-driven regime is investigated by cases with $\overline{\alpha_{\phi z}}=0$.

In reality, both magnetic tension and pressure contribute to driving outflows, and realistic MDWs from PPDs are somewhere between these two regimes.
In addition, the irradiation from a central star is expected to play a role in MDWs; such magneto-thermal winds have been intensively studied recently \citep{Bai2017a,2019ApJ...874...90W,2020ApJ...896..126G}.
The thermal contribution generally induces denser winds, and therefore, the relative contribution from magnetic tension is suppressed to give smaller $\overline{\alpha_{\phi z}}$.
Although our model adopt simple presctiptions for $\overline{\alpha_{\phi z}}$, in more realistic and elaborate models we have to carefully determine $\overline{\alpha_{\phi z}}$ with considering its spatial and temporal variation.
}

%%%%%%%%%%%%%%%%%%%%%%%%%%%%%%%%%%%%%%%%%%%%%%%%%%%%%%%%%%%%%%%%%%%%%%%
\subsection{Dust Surface Density}
\label{sect:dust}
The time evolution of dust surface density, $\Sigma_{\rm d}$, has
basically a similar form to eq.~(\ref{eq:gas}),
\begin{equation}
\label{eq:dust}
 \frac{\partial \Sigma_{\rm d}}{\partial t} +
 \frac{1}{r}\frac{\partial}{\partial r}(r\Sigma_{\rm d} v_{{\rm d},r})
 + \left(\rho_{\rm d} v_{{\rm d},z} \right)_{\rm w} = 0,
\end{equation}
where $ (\rho_{\rm g} v_{{\rm d},z} )_{\rm w} $ is the mass loss rate
of the solid component dragged upward by gaseous MDWs, 
$ (\rho_{\rm d} v_{{\rm d},z} )_{\rm w}  = D_{\rm w}  \left(\rho_{\rm
  d} c_{\rm s} \right)_{\rm mid} $.
The values of $v_{{\rm d}, r}$ and $D_{\rm w}$ depend on the masses and sizes of solid particles.
Under the single-size approximation, we use $v_{{\rm d}, r}$ and $D_{\rm w}$ for bodies with the mass-weighted averaged mass $m_{\rm p}$ (see derivation in Appendix \ref{app:single_size_approximation}).

\added{
In Eq.~(\ref{eq:dust}), we neglect the radial diffusion of the solid component by the drag force from turbulent gas.
When the dust grains are sufficiently small to be stirred by the gas turbulence, the variation of $\Sigma_{\rm d}$ is small within a scale height in our calculations.
Therefore the inward and outward mass fluxes by the turbulent diffusion are almost canceled out each other and the radial profile of $\Sigma_{\rm d}$ would not be affected by the turbulent diffusion.
On the other hand, if the radial variation of $\Sigma_{\rm d}$ is large, the turbulent diffusion has to be taken into account.
In our calculations, however, such a steep gradient of $\Sigma_{\rm d}$ appears only where solid particles grow to large bodies that are hardly affected by the gas drag.
}

We can obtain a fitting formula for $D_{\rm w}$ from the result of \citet{Miyake2016b}.
\begin{equation}\label{eq:Dw}
D_{\rm w} = \max \left(-1.8 {\rm St} + C_{\rm w} , 0\right) ,
\end{equation}
where ${\rm St}$ is the Stokes number for a spherical dust grain defined as \citep{Sato2016a}
\begin{equation}\label{eq:St}
{\rm St} = \frac{\pi \rho_{\rm m}a}{2 \Sigma_{\rm g}} \max \left(1, \frac{4a}{9 \lambda_{\rm mfp}}\right).
\end{equation}
Here $\rho _{\rm m} = 2$ g cm$^{-3}$ is the material density of the solid component, $a = (3 m_{\rm p}/4\pi \rho_{\rm m})^{1/3}$ is the radius of a solid particle with mass, $m_{\rm p}$, and $\lambda_{\rm mfp}$ is the mean free path between gas particles.
The mean free path is expressed as $\lambda_{\rm mfp} = \mu m_{\rm u}/\sigma_{\rm mol}\rho_{\rm g}$, where $\sigma_{\rm mol} = 2.0 \times 10^{-15} {\rm cm}^2$ is the collisional cross section between molecules.
When $a<(9/4)\lambda_{\rm mfp}$, the gas drag is in the Epstein 
regime, and otherwise if $a>(9/4)\lambda_{\rm mfp}$, it is in the
Stokes regime. When ${\rm St}\ll 1$, the solid particles are well coupled to
the gas. When ${\rm St}\gg 1$, the solid particles are hardly affected by
the gas drag, and their motion is almost independent of the gas flow.
Since we adopt $C_{\rm w}\sim 10^{-5}$, eq.~(\ref{eq:Dw}) indicates that only small dust grains that are tightly coupled to the gas with ${\rm St} < 10^{-5}$ are lost with MDWs and that the larger grains are left in the disk \citep[see][for the detail]{Miyake2016b}.

The radial drift velocity, $v_{{\rm d},r}$, is derived from the radial and azimuthal components of the equation of motion of solid particles \citep{Adachi1976a,Weidenschilling1977a,Takeuchi2002a},
\begin{equation}
v_{{\rm d},r}= \frac{v_{{\rm g},r} - 2 {\rm St} \eta v_{\rm k}}{1+ {\rm St}^{2}},
 \label{eq:vdr}
\end{equation}
where $\eta$ is a dimensionless pressure gradient force described below and $v_{\rm K}=r\Omega_{\rm K}$ is the Keplerian rotational velocity.
\added{We note that the radial gas velocity in Eq.~(\ref{eq:vdr}) is updated directly from Eq.~(\ref{eq:vgr}).}
From eq.~(\ref{eq:vdr}), the relative velocity, ${\mid v_{{\rm d},r} - v_{{\rm g},r}\mid}$, has a maximum speed at ${\rm St}=1$.

The dimensionless pressure gradient force $\eta$ is
\begin{equation}\label{eq:eta}
\eta = -\frac{1}{2 \rho_{\rm g} r \Omega_{\rm k}^{2}} \frac{\partial P}{\partial r},
\end{equation}
where $P=(\rho_{\rm g} c_{\rm s}^2)_{\rm mid} = \Omega_{\rm K}\Sigma_{\rm g} c_{\rm s}/\sqrt{2\pi}$
is gas pressure at the midplane.
Although $\eta$ is positive everywhere in the MMSN, MDWs locally change $\eta$ in, and in some cases, $\eta$ could take a negative value.

\subsection{Radial Drift Timescale}
\label{sect:rdtime}
From the radial velocity of solid particle, eq.~(\ref{eq:vdr}), the radial drift timescale, $t_{\rm drift}$, is defined as
\begin{empheq}[left={t_{\rm drift} \equiv \frac{r}{\mid v_{\rm d, r} \mid} \approx\empheqlbrace}]{alignat=2}
  \frac{r}{|v_{{\rm g},r}|} & \quad\left({\rm St} \ll \frac{{\mid v_{{\rm g},r}\mid}}{2{\mid \eta \mid} v_{\rm K}}\right),
  \label{eq:t_drift1}\\
  \frac{1+{\rm St}^2}{2{\mid \eta \mid}{\rm St}\Omega}  & \quad\left( {\rm St} \gg \frac{{\mid v_{{\rm g},r}\mid}}{2{\mid \eta \mid}v_{\rm K}} \right),
  \label{eq:t_drift2}
\end{empheq}
where eq.~(\ref{eq:t_drift1}) is for smaller particles that are strongly coupled to the gas and eq.~(\ref{eq:t_drift2}) is for larger particles that are weakly coupled to the gas. 
When the accretion is induced by turbulent viscosity, $\frac{{\mid v_{{\rm g},r}\mid}}{2{\mid \eta \mid}v_{\rm K}}\approx \frac{\overline{\alpha_{r\phi}}c_{\rm s}^2}{2\eta v_{\rm K}^2}$. 
If we consider typical conditions of the MMSN, $c_{\rm s}^2/v_{\rm K}^2\sim 10^{-3}$ and $\eta \sim 10^{-3}$,
we get $\frac{{\mid v_{{\rm g},r}\mid}}{2 {\mid \eta \mid}v_{\rm K}} \sim \overline{\alpha_{r\phi}}$; when solid particles grow to St$\gtrsim \overline{\alpha_{r\phi}}$, they enter the loose coupling state from the strong coupling limit.

We should note, however, that the above estimate is modified when the MDW torque plays a dominant role in determining $v_{{\rm g},r}$.
In addition, $\eta$ is largely altered in PPDs with MDWs, which we examine in Section \ref{results}.

%%%%%%%%%%%%%%%%%%%%%%%%%%%%%%%%%%%%%%%%%%%%%%%%%%%%%%%%%%%%%%%%%%%%%%%
\subsection{Collisional Growth of Solid Particles}\label{Sect:growth}
We assume perfect sticking via collisions without collisional
fragmentation and adopt a single-size approximation to
calculate the collisional growth of dust particles,
following \citet{Sato2016a}.
The time evolution of $m_{\rm p}$ via collisions and radial drift 
is given by (see derivation in Appendix \ref{app:single_size_approximation}),
\begin{equation}
\label{eq:growth}
 \frac{\partial m_{\rm p}}{\partial t} + v_{{\rm d},r}\frac{\partial
   m_{\rm p}}{\partial r} = \frac{2 \sqrt{\pi} a^{2} \Delta v_{\rm
     pp}}{h_{\rm d}} \Sigma_{\rm d},
\end{equation} 
where $\Delta v_{\rm pp}$ is the relative velocity between solid
particles with masses $\sim m_{\rm p}$ and $h_{\rm d}$ is the scale
height of dust grains with $m_{\rm p}$,
which can be estimated as
\begin{equation}\label{eq:hd}
h_{\rm d} = h_{\rm g} \left( 1+ \frac{{\rm St}}{\overline{\alpha}} \frac{1+2{\rm St}}{1+{\rm St}} \right)^{-\frac{1}{2}},
\end{equation}
\citep{Youdin2007b}, where $\overline{\alpha}$ is turbulent strength
and we here adopt $\overline{\alpha} = \overline{\alpha_{r\phi}}$.
We follow the growth of $m_{\rm p}$ up to $a=1$ km, because we are interested in the formation processes of km-sized bodies, planetesimals, and the growth process of bodies larger than $\sim 1$ km is controlled by the self-gravity,  which is not considered in our model.

We basically follow \citet{Okuzumi2012a} for the derivation of $\Delta v_{\rm pp}$, which we briefly summarize below:
\begin{equation}
\label{eq:delta_vpp}
 \Delta v_{\rm pp} = \sqrt{ \Delta v_{\rm B}^2 +  \Delta v_{r}^2 +  \Delta v_{\phi}^2 + \Delta v_{z}^2 + \Delta v_{\rm t}^2}.
 \end{equation}
$\Delta v_{\rm B}$ denotes the relative velocity by the Brownian motion, which is given by
 \begin{equation}
 \Delta v_{\rm B} = \sqrt{\frac{8(m_1+m_2)k_{\rm B} T}{\pi m_1 m_2}},
 \end{equation}
where $m_1$ and $m_2$ are the masses of colliding particles, and in
the single-size approximation we take $m_1=m_2=m_{\rm p}$.
 $\Delta v_{r}$, $\Delta v_{\phi}$, and $\Delta v_z$ are the three
components of the relative velocity between colliding particles, which
arise from the drift motion of dust grains from the background gas.
In addition to the $r$ component of eq.~(\ref{eq:vdr}), the $\phi$ and $z$
components are written as
\begin{equation}
  v_{{\rm d},\phi} = - \frac{\eta v_{\rm k}}{1 + {\rm St}^2}
\end{equation}
and
\begin{equation}
  v_{{\rm d},z} = - \frac{{\rm St} \Omega_{\rm k}}{1+{\rm St}} z.
\end{equation}
If we apply the single-size approximation ($m_1=m_2=m_{\rm p}$) in a strict sense,  $\Delta v_{r}$, $\Delta v_{\phi}$, and $\Delta v_z$ are all zero.
However, this is not realistic.
We estimate them by $\Delta v_{{\rm d},i} = | v_{{\rm d},i}({\rm St}_1) - v_{{\rm d}, i}({\rm St}_2) |$, where ${\rm St}_1={\rm St} $, ${\rm St}_2=\epsilon {\rm St}$.
In \citet{Sato2016a}, they found that $\epsilon = 0.5$ is a good approximation that well explains the result considering the realistic distribution of particle masses by \citet{Okuzumi2012a}.
\added{We employ $\epsilon = 0.5$ in our calculations, following these works.}

The $z$ component can be estimated from the difference between the vertical sedimentation velocities of particles 1 and 2 as
\begin{eqnarray}
\label{eq:dvz}
  \Delta v_{z} = \frac{1}{\sqrt{2\pi} h_{{\rm d},12}}
  \int^{\infty}_{-\infty} \left| v_{{\rm d},z}({\rm St}_1) - v_{{\rm
      d},z}({\rm St}_2) \right| \nonumber \\
  \times \exp \left(-\frac{z^2}{2 h_{{\rm d},12}^2} \right) dz, \nonumber \\
\added{  = \sqrt{\frac{2}{\pi}} \left| \frac{{\rm St}_{1}}{1+{\rm St}_{1}} - \frac{{\rm St}_{2}}{1+{\rm St}_{2}} \right|h_{{\rm d}, 12} \Omega_{\rm k},}
 \end{eqnarray}
where $h_{\rm d,12} = (h_{\rm d,1}^{-2} + h_{\rm d,2}^{-2})^{\added{-}1/2}$ is the mean scale height of particles with $m=m_1$ and $m=m_2$.
Here, $h_{{\rm d},i} (i=1,2) $ stands for $h_{\rm d}({\rm St}_i)$.
\added{
The integral of eq.~(\ref{eq:dvz}) is important for $|z| < h_{\rm d}$ where ${\rm St}$ and $\Omega_{\rm k}$ are constant on ${\cal O}(z)$, so that the second equality of eq.~(\ref{eq:dvz}) is approximately obtained.
}

$\Delta v_{\rm t}$ is the relative velocity by the turbulent motion of the gas that pushes particles in a stochastic manner.
We adopt the formula introduced in \citet{Ormel2007a}, in which they classified the interaction between turbulent eddies and particles into two regimes.
In Class I solid particles interact with an eddy for a long time and their initial motion is modified before the eddy vanishes;
In Class $\rm I\hspace{-.1em}I$, the duration of the interaction between solid particles and an eddy turnover time is shorter than their stopping time.
We define the relative velocity of each class as $\Delta v_{\rm I}$ and $\Delta v_{\rm I\hspace{-.1em}I}$, respectively.
From these arguments, the relative velocity can be estimated as
\begin{equation}
\label{t*case}
\Delta v_{\rm t}^2 = \left\{\begin{array}{l}
\Delta v_{\rm I}^2 \ ( 1.6 {\rm St}_1  < Re_{\rm t}^{-1/2}) \\
\Delta v_{\rm I}^2+\Delta v_{\rm I\hspace{-.1em}I}^2 \ (Re_{\rm t}^{-1/2} < 1.6{\rm St}_1 < 1.0) \\
\Delta v_{\rm I\hspace{-.1em}I}^2 \ (1.0 < 1.6{\rm St}_1).
\end{array} \right.
\end{equation}
$Re_{\rm t} = \overline{\alpha} c_{\rm s} h_{\rm g} /\nu_{\rm m}$ is a turbulent Reynolds number, where $\nu_{\rm m} = \lambda_{\rm mfp} v_{\rm th}/2$ is a kinematic viscosity of the gas.
The thermal velocity is given by $v_{\rm th} = \sqrt{8/\pi} c_{\rm s}$.
Here
\begin{equation}
  \Delta v_{\rm I}^2 = c_{\rm s}^2\overline{\alpha}
  \frac{{\rm St}_2-{\rm St}_1}{{\rm St}_2+{\rm St}_1}  \left[ \frac{{\rm St}_1^2}{{\rm St}_1 +1} -
    \frac{{\rm St}_1^2}{{\rm St}_1+{\rm St}_{12}^* }  + (1 \leftrightarrow 2)\right], 
 \label{eq:vI}
\end{equation}
and
\begin{eqnarray}
%%%
\Delta v_{\rm I\hspace{-.1em}I}^2 = c_{\rm s}^2\overline{\alpha}
 \left[ ({\rm St}_{12}^* - Re_{\rm t}^{-1/2}) \right.
 + \frac{{\rm St}_1^2}{{\rm St}_1 + {\rm St}_{12}^*} \nonumber \\
 \left. - \frac{{\rm St}_1^2}{{\rm St}_1+Re_{\rm t}^{-1/2}} + (1 \leftrightarrow 2)\right],
\label{eq:vII}
\end{eqnarray}
where ${\rm St}_{12}^* =  \max[Re_{\rm t}^{-1/2}, \min(1.6 {\rm St}_1, 1)]$ is the
boundary between the two regimes and $(1 \leftrightarrow 2)$ denotes
interchange between particles 1 and 2.

\subsection{Growth Timescale}
We define a growth timescale, $t_{\rm grow}$, of solid particles from the increase of particle size, $a$;
we here use $a$, instead of $m_{\rm p}$, to evaluate $t_{\rm grow}$ because it coincides with the growth timescale of St in the Epstein regime (eq.~\ref{eq:St}). 
Then, from eq.~(\ref{eq:growth}) we can derive 
\begin{eqnarray}
  \label{eq:t_grow1}
  t_{\rm grow} &\equiv& \left(\frac{1}{a}\frac{\mathrm{d}a}{\mathrm{d}t}\right)^{-1}
  = \left(\frac{1}{3m_{\rm p}}\frac{\mathrm{d}m_{\rm p}}{\mathrm{d}t}\right)^{-1}\nonumber \\
  &=&  \frac{3m_{\rm p}h_{\rm d}}{2\sqrt{\pi}a^2 \Delta v_{\rm pp} \Sigma_{\rm d}}.
\end{eqnarray}
In this paper, we mainly focus on the dust growth in a range of $\overline{\alpha_{r\phi}} \ll {\rm St} \ll 1$.
In this case, $h_{\rm d} \approx h_{\rm g}(1+{\rm St}/\overline{\alpha_{r\phi}})^{-1/2}
\approx h_{\rm g} \sqrt{\overline{\alpha_{r\phi}}/{\rm St}}$ in eq.~(\ref{eq:hd}).
$\Delta v_{\rm pp}$ is dominated by the turbulent component in usual situations.
The regime of $\overline{\alpha_{r\phi}} \ll {\rm St} \ll 1$ roughly corresponds to the intermediate range of eq.~(\ref{t*case}), which gives $\Delta v_{\rm pp} \approx \sqrt{\overline{\alpha_{r\phi}} {\rm St}}c_{\rm s}$.
Then, we finally obtain the following approximated expression of $t_{\rm grow}$ for $\overline{\alpha_{r\phi}} \ll {\rm St} \ll 1$:
\begin{eqnarray}
  \label{eq:t_grow2}
  t_{\rm grow}&\approx& \frac{4{\rm St}}{\sqrt{\pi}}\frac{\Sigma_{\rm g}}{\Sigma_{\rm d}}\min\left(1,\frac{9\lambda_{\rm mfp}}{4a}\right)\frac{h_{\rm g}\sqrt{\overline{\alpha_{r\phi}}/{\rm St}}}{\sqrt{\overline{\alpha_{r\phi}} {\rm St}}c_{\rm s}} \nonumber \\
  &=& \frac{4}{\sqrt{\pi}\Omega_{\rm K}}\frac{\Sigma_{\rm g}}{\Sigma_{\rm d}}\min\left(1,\frac{9\lambda_{\rm mfp}}{4a}\right).
\end{eqnarray}

\subsection{Initial Conditions and Numerical Methods}
\label{numerical}
We adopt the MMSN model for the initial gas surface density,
\begin{equation}
\label{MMSN}
  \Sigma_{\rm g} = \Sigma_{\rm g,0} \left( \frac{r}{1 \rm
    au}\right)^{-\frac{3}{2}},
\end{equation}
where $\Sigma_{\rm g,0} = 1700\ {\rm g \ cm^{-2}}$ at $r=1$ au is taken from the original value introduced in \citet{Hayashi1981a}.
The initial dust surface density is  $\Sigma_{{\rm d},0} = \Sigma_{{\rm g},0}/100 $, which is adopted from the interstellar dust-gas ratio.
We follow the collisional growth of dust grains that have the initial radius of $a = 1 \times 10^{-4} \rm{cm}$, which is the same order of magnitude as the typical size of interstellar dust grains.

%%%%%%%%%%%%%%%%%%%%%%%%%%%%%%%%%%%%%%%%%%%%%%%%%%%%%%%%%%%%%%%%%%%%%%%
The simulation domain covers a region from 0.01 au to 300 au, which is resolved by 255 grid points\footnote{\added{We checked the numerical convergence by using the calculation with up to 8 times higher resolution.}}.
The grid spacing, $\Delta r$, is proportional to $\log (r)$.
The surface density of the gas is updated by solving eq.~(\ref{eq:gas}) in a time-implicit manner.
The initial total mass of the gas component is $\approx 0.04M_{\odot}$.
\added{
At the inner and outer boundaries, the only  outflow of the mass is allowed.
When the mass is outgoing from the simulation box at the boundaries ($v_{{\rm g},r}\le 0$ at $r=r_{\rm in}$; $v_{{\rm g},r}\ge 0$ at $r=r_{\rm out}$), the non-gradient condition is allplied to $v_{{\rm g},r}$ and $\Sigma_{\rm g}$; when the mass is incoming at the boudary, $v_{{\rm g},r}$ is set to be 0 there.
We note that the treatment of the inner boundary condition for the gas component does not affect the evolution of the solid component, because, as we explain later, we adopt $\Sigma_{\rm d}=0$ in $r<1.5$ au, which is a sufficiently outer location from the inner boundary of $r=0.01$ au.
}

We solve eqs.~(\ref{eq:dust}) and (\ref{eq:growth}) to update the surface density of the solid component and the peak-mass \deleted{in a time-explicit manner}.
\added{
These equations are solved explicitly in time and space by using the first-order upwind scheme.
The reason why we choose the different method from the gas component is that the timescale of dust growth is short and the dust evolution should be resolved with much smaller time steps.
Since we update the dust component in an explicit manner, we determine the time step to ensure the numerical stability of both $\Sigma_{\rm d}$ and $m_{\rm p}$.
On the other hand, we set a longer fixed time step for the gas component that adopts the implicit time update.
}

We consider icy dust grains in this paper and assume $\Sigma_{\rm d}=0$ inside $r < 1.5$ au throughout the time evolution, whereas we do not treat detailed properties of the snowline \citep{2011ApJ...738..141O} for simplicity.
In addition, unless we do not take into account the fragmentation of collisionally grown particles, $m_{\rm  p} $ rapidly increases because of the accelerated coagulation in $r\lesssim 1$ au in an unrealistic manner \citep{Birnstiel2010a, Sato2016a}.
This is also a reason why we set $\Sigma_{\rm d}=0$ in $r<1.5$ au to avoid unphysically rapid growth of dust grains.

We also assume $\Sigma_{\rm d}=0$ outside $r > 200$ au to avoid the effect of the outer boundary on the surface density profile of the dust component.
The initial total mass of the dust component is $\approx 103M_{\rm E}$, where $M_{\rm E}$ is the Earth mass.

We calculate the time evolution of $\Sigma_{\rm g}$, $\Sigma_{\rm d}$, and $m_{\rm p}$ for different sets of the 3 parameters, the turbulent viscosity, $\overline{\alpha_{r\phi}}$, the mass loss by the MDW, $C_{\rm w}$, and the MDW torque, $\overline{\alpha_{\phi z}}$.
The employed values are summarized in Table \ref{parameter}.

\begin{deluxetable*}{cccc|c|c}
\tablecaption{
  List of adopted parameters.
  The second right column shows the result of the dust growth to St $>1$. 
  The right column indicates the corresponding figure number.
  We use following abbreviations: STb $=$ strong turbulence, WTb $=$ weak turbulence, NW $=$ no wind mass loss, WM $=$ with wind mass loss, ZTq $=$ zero-torque, CTq $=$ constant torque, and STq $=$ $\Sigma$-dependent torque.
\label{parameter}}
\tablehead{ Case & $\overline{\alpha_{r \phi}}$ & $C_{\rm W}$ & $\overline{\alpha_{\phi z, 0}}$ & Growth ? & Corresponding Figure}
\startdata
STb + NM + ZTq  &$8.0\times 10^{-3}$&$0.0$              &0.0                 &  No & \ref{aND3}   \\
STb + WM + ZTq  &$8.0\times 10^{-3}$&$1.0\times 10^{-5}$&0.0                 &  No & \ref{aD3}    \\
WTb + NM + ZTq  &$8.0\times 10^{-5}$&$0.0$              &0.0                 &  No & - \\
WTb + NM + CTq  &$8.0\times 10^{-5}$&$0.0$              &$1.0\times 10^{-4}$ &  No & - \\
WTb + NM + STq  &$8.0\times 10^{-5}$&$0.0$              &$1.0\times 10^{-5}$ &  No & - \\
WTb + WM + ZTq  &$8.0\times 10^{-5}$&$1.0\times 10^{-5}$&0.0                 & Yes & \ref{inaD3}  \\
WTb + WM + CTq  &$8.0\times 10^{-5}$&$1.0\times 10^{-5}$&$1.0\times 10^{-4}$ &  No & \ref{inaDc3} \\
WTb + WM + STq  &$8.0\times 10^{-5}$&$1.0\times 10^{-5}$&$1.0\times 10^{-5}$ & Yes & \ref{inaDsw3}\\
\enddata
\end{deluxetable*}
 %%%%%%%%%%%%%%%%%%%%%%%%%%%%%%%%%%%%%%%%%%%%%%%%%%%%%%%%%%%%%%%%%%%%%%%

\section{Results} \label{results}
\subsection{MRI active ($\overline{\alpha_{r\phi}}=8\times 10^{-3}$) cases}
\label{MDWeffect}

\begin{figure}
  \centering
  \includegraphics[keepaspectratio, scale=0.8, angle=0]{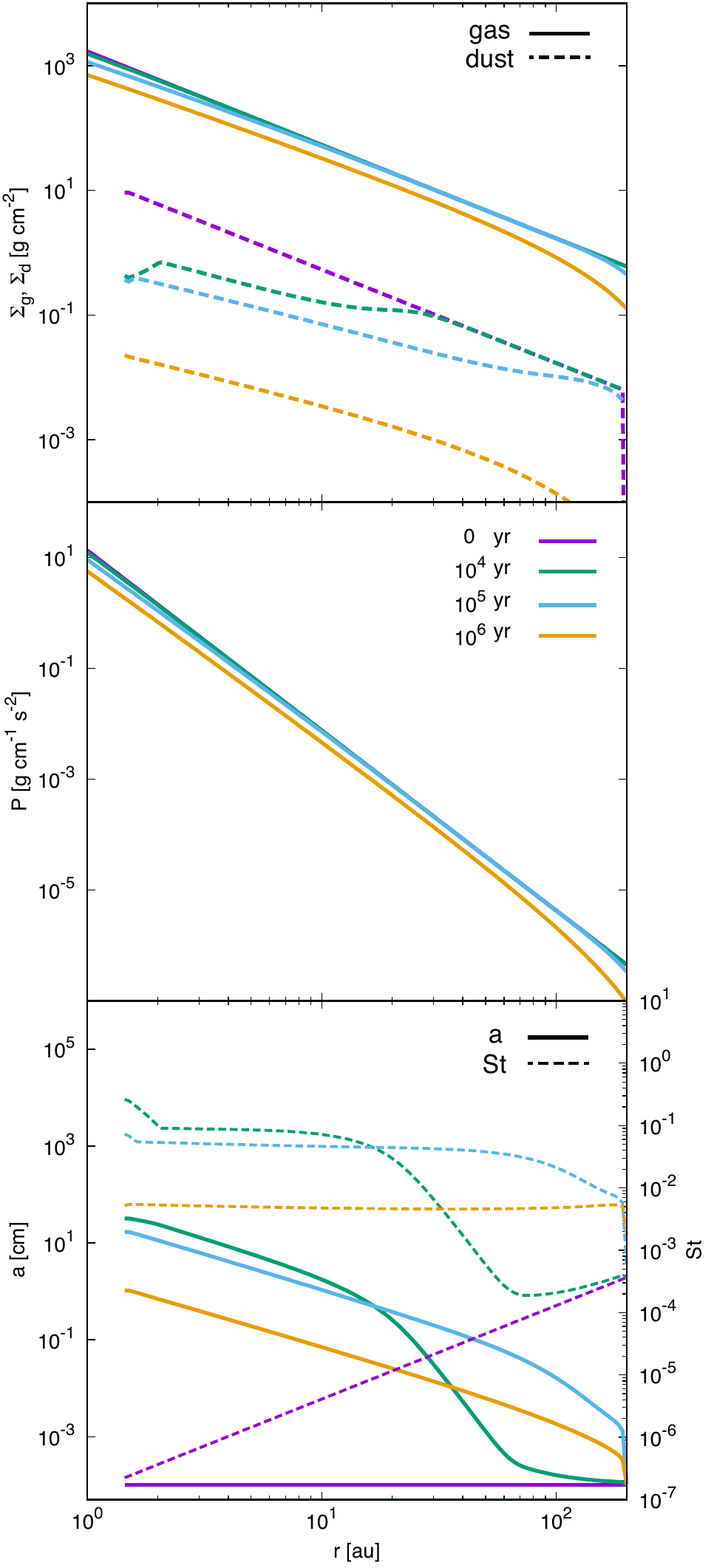}
  \caption{
Time evolution of the gas (solid) and dust (dashed) surface densities ({\it top}), gas pressure at the midplane ({\it middle}), and the radius (solid) and Stokes number (dashed) of solid particles ({\it bottom}) of the case with strong turbulence + no mass loss + zero-torque (STb+NM+ZTq).
Each colored line denotes the snapshot at $t=$ 0 (purple), $10^4$ yrs (green), $10^5$ yrs (light blue), and $10^{6}$ yrs (orange).
}
\label{aND3}
\end{figure}

In this section, we present results of the MRI active cases with $\overline{\alpha_{r\phi}}=8\times 10^{-3}$.
Figure~\ref{aND3} compares the radial profiles of
of various physical quantities of the case without MDW (STb+NM+ZTq) at $t=0, 10^{4}, 10^{5},$ and $10^{6}$ yrs.
The top panel shows that the surface density of the gas, $\Sigma_{\rm g}$, decreases with time by viscous accretion.
The gas pressure at the midplane, $P$, (middle panel) as well as $\Sigma_{\rm g}$ keeps a decreasing trend with $r$, which leads to the inward drift of solid particles (eqs.~\ref{eq:vdr} \& \ref{eq:eta}).

The decrease of the dust surface density, $\Sigma_{\rm d}$, is more significant in an inside-out manner (top panel of Figure~\ref{aND3}); $\Sigma_{\rm d}$ is excavated in $r \lesssim 30$ au at $t=10^{4}$ yrs and $\lesssim 150$ au at $10^{5}$ yr, respectively.
This is because dust grains at smaller $r$ reach the drift limited state from earlier times.
The dust grains are initially well coupled to the gas because St $<10^{-3}$ in the entire region (bottom panel of Figure~\ref{aND3}), and therefore $t_{\rm grow} < t_{\rm drift}$.
The growth of dust grains is more rapid in the inner region because $t_{\rm grow} \propto \Omega_{\rm K}^{-1}$ (eq.~\ref{eq:t_grow2}).
The increase of St with the dust growth reduces $t_{\rm drift}$ ($\propto {\rm St}^{-1}$ for St $\ll 1$; eq.~\ref{eq:t_drift2}).
Eventually $t_{\rm drift}\approx t_{\rm grow}$ at St $\approx 0.1$, and finally, dust particles drift inward rapidly before growing to further larger bodies; the growth of solid particles are limited by the radial drift \citep[bottom panel of Figure \ref{aND3}; ][see also Section~\ref{sect:rdrift}]{Okuzumi2012a, Sato2016a}.

Readers may notice an upward bend of St near the inner edge of the dust disk at $t=10^{4}$ and $10^{5}$ yrs.
Inside this point the solid particles are in the Stokes regime.
However, they are still subject to the inward drift and do not overcome the radial drift barrier.

\begin{figure}
  \centering
  \includegraphics[keepaspectratio, scale=0.8, angle=0]{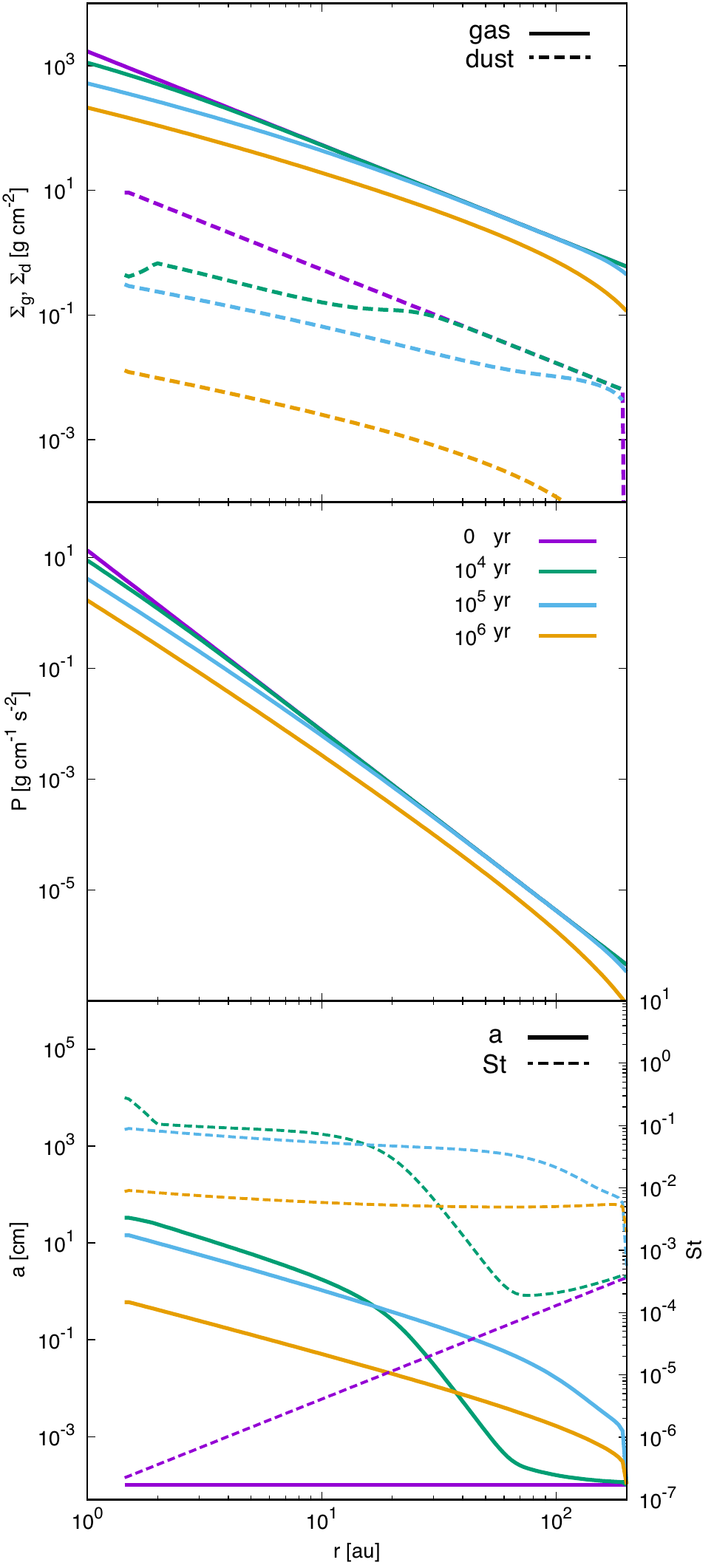}
  \caption{
Same as Figure~\ref{aND3} but for the case with strong turbulence + weak mass loss + zero-torque (STb+WM+ZTq).
}
\label{aD3}
\end{figure}

Figure~\ref{aD3} shows the case with the mass loss by (STb+WM+ZTq).
The top panel shows that $\Sigma_{\rm g}$ decreases slightly faster particularly in the inner region by the MDW than $\Sigma_{\rm g}$ of the case without MDWs (Figure~\ref{aND3}).
However, the difference of $\Sigma_{\rm g}$ between the two cases is not so significant because the evolution of the gas is mainly controlled by the strong accretion owing to the large $\overline{\alpha_{r\phi}}$.

Because the time evolution of $\Sigma_{\rm g}$ of Figure~\ref{aND3} is similar to that of Figure~\ref{aD3}, the properties of the solid component also follow the similar trend.
The mass loss of solid particles by the gaseous MDWs is not effective in both cases.
Substituting $C_{\rm w}=2\times 10^{-5}$ into eq.~(\ref{eq:Dw}) shows that dust grains should be tightly coupled with ${\rm St} \lesssim 10^{-5}$ at the midplane in order to be entrained by the MDW.
In the inner region where the effect of the MDWs is significant dust grains grow rapidly beyond ${\rm St} > 10^{-5}$.
Therefore, most of the solid particles are not lost with the MDWs but left in the disk and eventually accrete to the central star.

The results presented in Figures \ref{aND3} \& \ref{aD3} indicate that the collisional growth cannot proceed to form planetesimals in the MRI-active condition with $\overline{\alpha_{r\phi}} = 8\times 10^{-3}$, whether or not the gas is lost via MDWs.
This is because the evolution of the gas component is mainly determined by the accretion. 
As a result, the solid particles inevitably drift inward when they grow to ${\rm St}\approx 0.1$.
Although in these cases we do not consider the MDW torque, it does not affect so much the overall evolution of $\Sigma_{\rm g}$ because the accretion is induced dominantly by the turbulent viscosity.
Smaller $\overline{\alpha_{r\phi}}$ is required to overcome the radial drift barrier.


%%%%%%%%%%%%%%%%%%%%%%%%%%%%%%%%%%%%%%%%%%%%%%%%%%%%%%%%%%%%%%%%%%%%%%%%
\subsection{MRI inactive ($\overline{\alpha_{r\phi}}=8\times 10^{-5}$) cases }
\label{dustring}

\begin{figure}
  \centering
  \includegraphics[keepaspectratio, scale=0.8, angle=0]{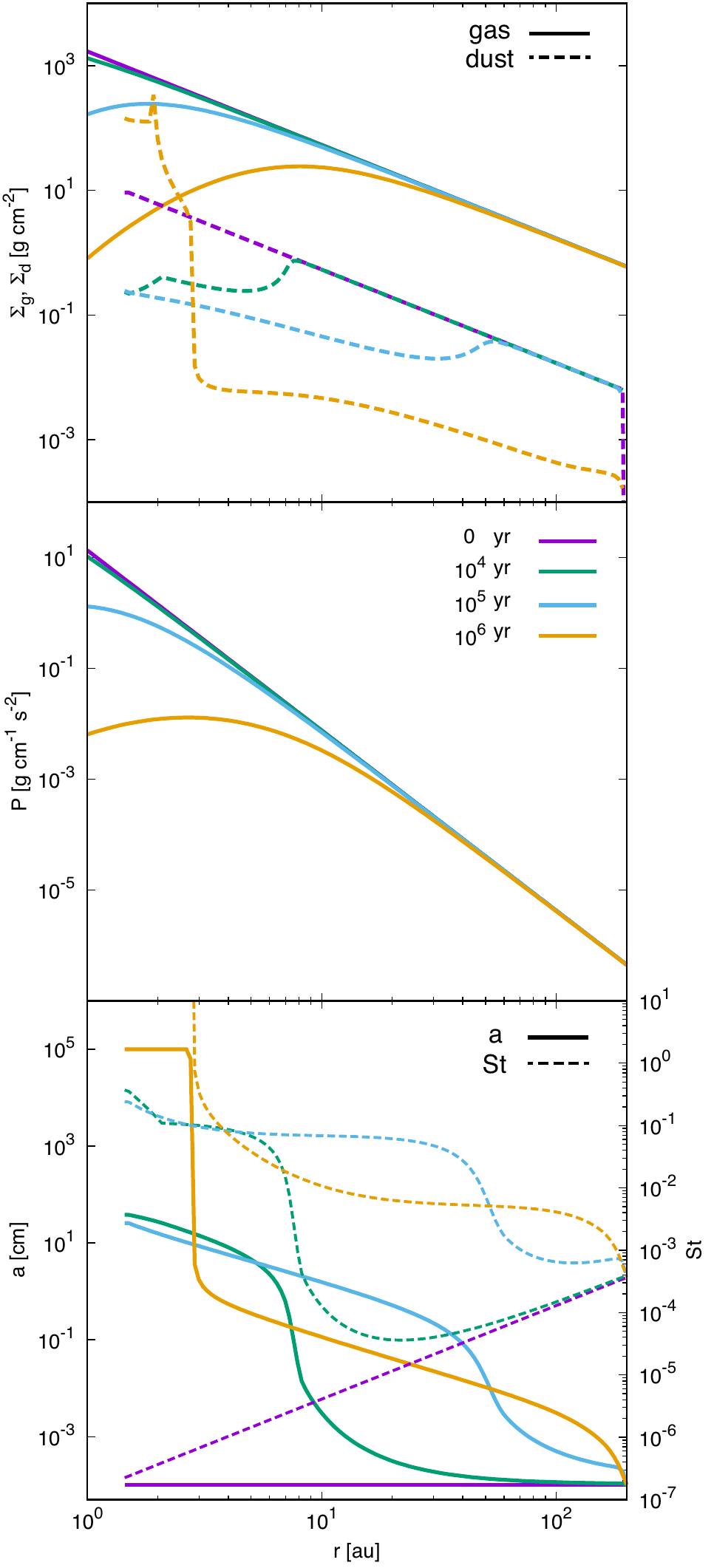}
  \caption{
Same as Figure~\ref{aND3} but for the case with weak turbulence + weak mass loss + zero-torque (WTb+WM+ZTq).
}
\label{inaD3}
\end{figure}

\begin{figure}
  \centering
  \includegraphics[keepaspectratio, scale=0.8, angle=0]{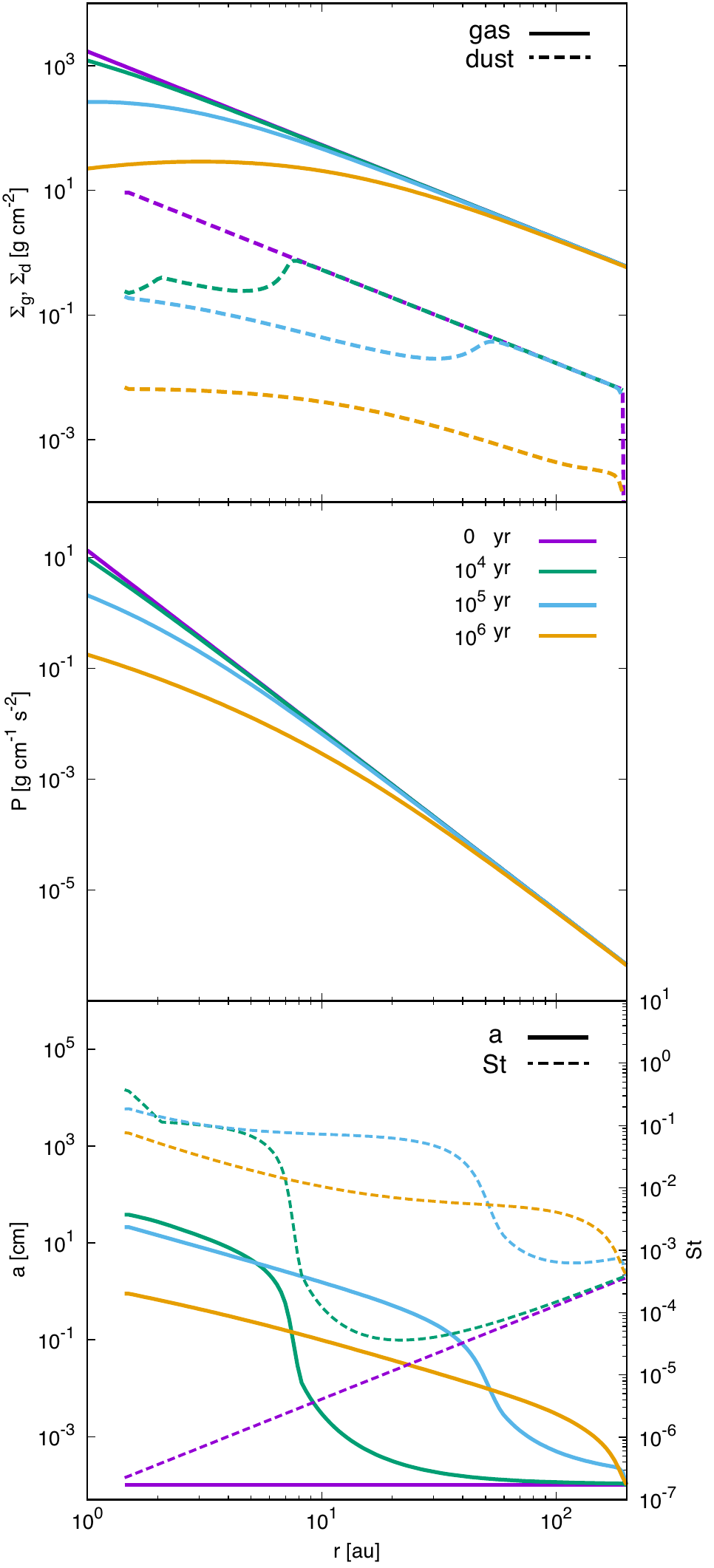}
  \caption{
Same as Figure~\ref{aND3} but for the case with weak turbulence + weak mass loss + constant torque (WTb+WM+CTq).
}
\label{inaDc3}
\end{figure}

\begin{figure}
  \centering
  \includegraphics[keepaspectratio, scale=0.8, angle=0]{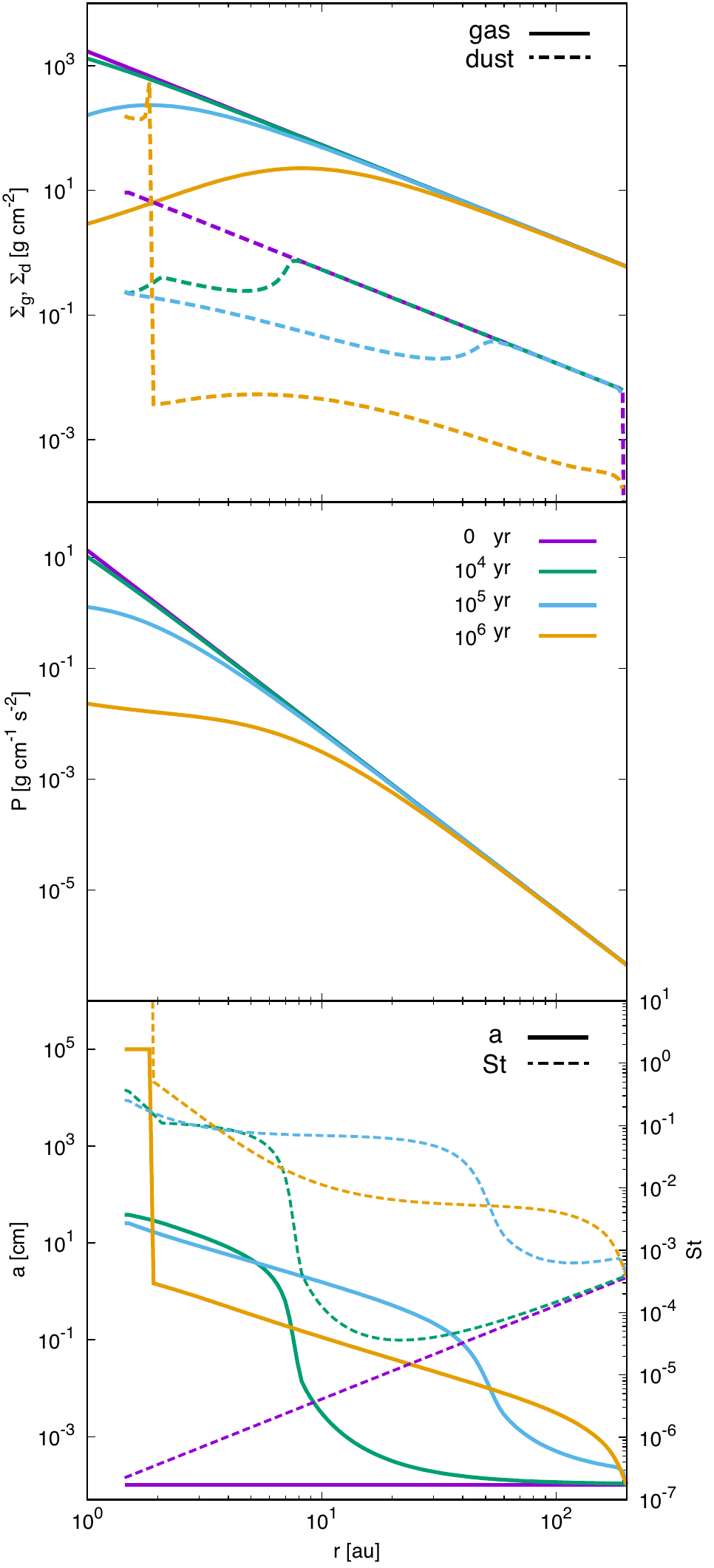}
  \caption{
Same as Figure~\ref{aND3} but for the case with weak turbulence + weak mass loss + $\Sigma$-dependent torque (WTb+WM+STq).
}
\label{inaDsw3}
\end{figure}

In this subsection, we examine the time evolution of MRI inactive cases with MDWs.
We focus particularly on the evolution of the three cases with the mass loss by MDWs (WTb+WM+XTq cases in Table \ref{parameter}) that adopt different models of the wind torque (zero / constant / $\Sigma$-dependent torque).
In addition to these 3 cases, we also performed cases without mass loss by MDWs (WTb+NM+XTq in Table \ref{parameter}).
They are not displayed in figures because most of the solid particles fall into the central star by the radial drift, similarly to the MRI active cases.

Figure~\ref{inaD3} presents the results of the case with weak turbulence + wind mass loss + zero-torque (WTb+WM+ZTq).
The top panel shows that the mass loss by the MDW plays an effective role; the radial profile of $\Sigma_{\rm g}$ largely deviates from the initial power-law profile in $r \lesssim 3$ (10) au at $t=10^{5}$ ($10^{6}$) yrs.

The top panel of Figure~\ref{inaD3} also shows that although the time evolution of $\Sigma_{\rm d}$ up to \replaced{$10^{5}$}{$t=10^{5}$} yrs is similar to that of the strong turbulent cases (Figures \ref{aND3} and \ref{aD3}), there is a region where dust particles are piled up near the inner edge ($r=1.5$ au) of the dust disk at \replaced{$10^{6}$}{$t=10^{6}$} yrs.

The bottom panel of Figure~\ref{inaD3} shows that the size of the dust particles grows up to 1 km, which is the upper cap in our setting, at \replaced{$t=1$ Myr}{$t=10^{6}$ yr} in a ring-like region of $\Sigma_{\rm d}$ in $r\lesssim 3$ au (top panel).
St goes beyond $>1$ at that time, which is in contrast to St $\approx 0.1$ at the earlier times of \replaced{$t=0.01$ and 0.1 Myr}{$t=10^{4}$ and $10^{5}$ yr} when the growth of solid particles are still constrained by the radial drift.
We can conclude that dust particles get over the radial drift barrier in this case.
A key to the dust growth against the radial drift barrier is to achieve St $>1$, which we explain in more detail later (Section \ref{sec:mechanism-rapid-dust}).

In the constant torque case (WTb+WM+CTq), the mass accretion by the turbulent viscosity is as weak as the zero-torque case.
However, the top panel of Figure~\ref{inaDc3} shows that $\Sigma_{\rm g}$ in the inner region is not so low as that obtained in Figure~\ref{inaD3} because the gas is supplied by the wind-driven accretion.
As a result, the gas pressure does not show a local maximum but monotonically decreases with $r$. Dust grains suffer inward radial drift before growing to St $\gtrsim 1$.

In the $\Sigma$-dependent torque case (WTb+WM+STq), the evolution of $\Sigma_{\rm g}$ exhibits intermediate behavior between the previous two cases as shown in the top panel of Figure~\ref{inaDsw3}.
$\Sigma_{\rm d}$ (top panel) shows a ring-like structure, and the solid component grows beyond St $>1$ to reach the upper bound of $a=1$ km (bottom panel).
However, contrary to the zero-torque case, this case does not show a local maximum  of the gas pressure (middle panel); {\it a pressure bump is not a necessary condition to form a local concentration of dust grains} and the subsequent growth of solid particles.
This issue is explained in more detail in Section \ref{sec:mechanism-rapid-dust}.

A difference of $\Sigma_{\rm d}$ between the ZTq and STq cases (Figures~\ref{inaD3} and \ref{inaDsw3}) is the width of the ring-like region of the solid component.
The top panel of Figure~\ref{inaD3} shows a wider ring with high $\Sigma_{\rm d}$  at \replaced{$t=1$ Myr}{$t=10^{6}$ yr} because the pressure maximum \added{where the drift of dust grains are halted,} moves outward.
\added{
We note that the direction of the radial drift of the ZTq case (Figure~\ref{inaD3}) is outward while it is inward in the STq case (Figure~\ref{inaDsw3}).
Although the drift directions could be an additional explanation to the difference of the ring widths between the two cases, the contribution is almost negligible because the radial drift velocity of solid particles in the ring-like regions is too slow, $|v_{{\rm d},r}|\lesssim 10^{-9}$ au yr$^{-1}$, to modify their profiles.
Once dust particles enter the ring-like region from the outer part, they rapidly grow up to $a=1$ km ($=$ the maximum size in our setup) to remain there during the disk lifetime.
}
\deleted{Since the dust particles drift outward in the inner side of the pressure bump, the ring with high $\Sigma_{\rm d}$ expands outward.
As it will be discussed in Section~\ref{sec:backreaction-gas}, however, this outward migration of grown dust particles could be suppressed if an effect of the backreaction from dust to gas is considered.}

\subsection{Conditions for Dust Growth}
\label{sec:mechanism-rapid-dust}

\begin{figure*}
   \centering
  \includegraphics[keepaspectratio, scale=1.0, angle=0]{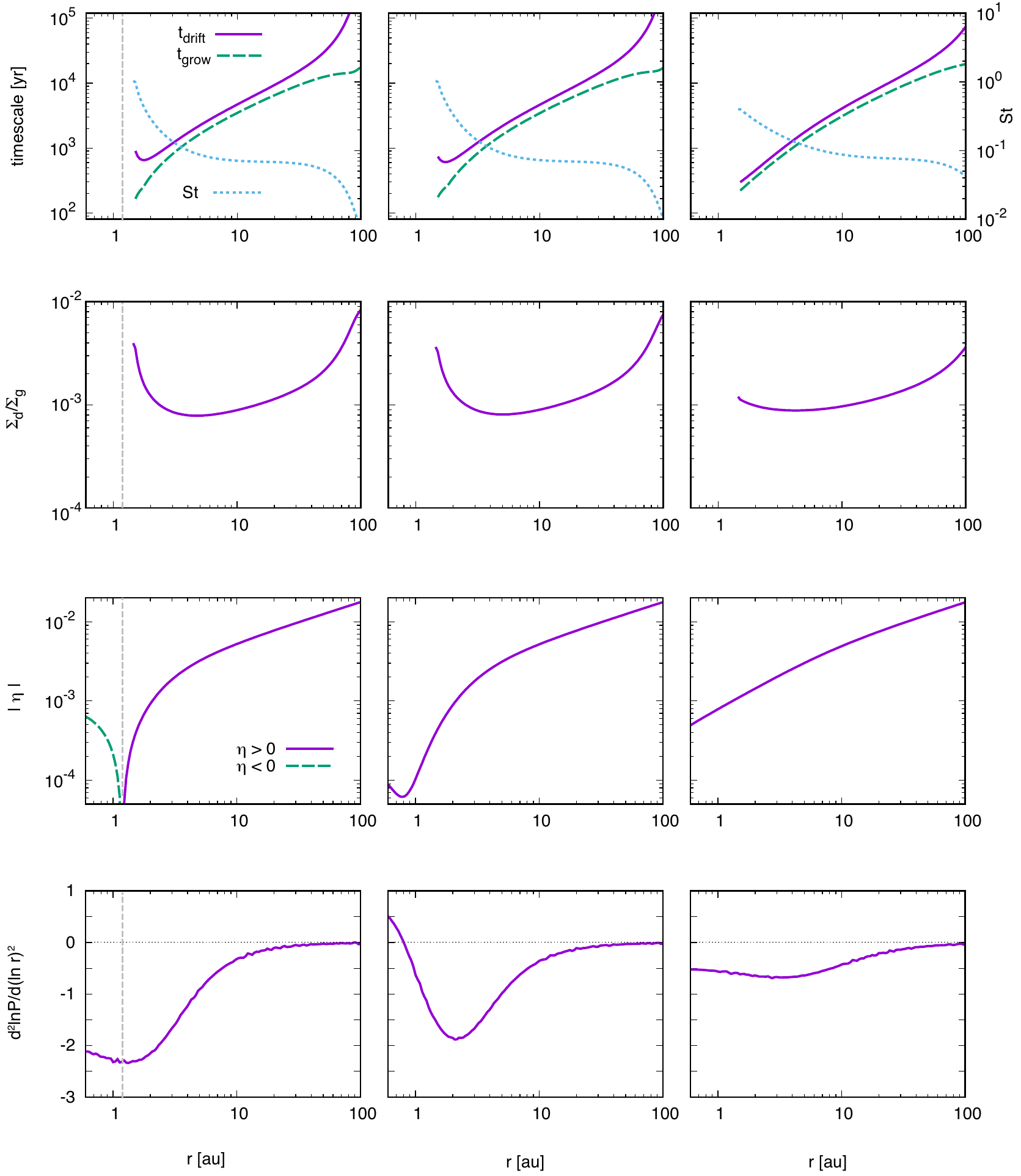}
  \caption{
Radial profile of following quantities for the case with WTb+WM+ZTq (left column), WTb+WM+STq (middle column), and WTb+WM+CTq (right column):
In the left (middle) column, all panels show snapshots at $t=1.77\times 10^{5}$ yrs ($t=1.86\times 10^{5}$ yrs) just when the Stokes number reaches unity at $r=1.5$ au.
In the right column, all panels show snapshots at $t=2.5\times10^5$ yrs when the dust evolution reaches a quasi-steady state.
({\it top row}): Drift (purple solid line) and growth (green dashed line) timescale of dust particles (left axis), and Stokes number of dust particles (blue-dotted line; right axis).
({\it second row}): Dust-to-gas surface density ratio.
({\it third row}): Absolute value of $\eta$.
The purple solid (green dashed) line represents the region with $\eta > (<) 0$.
The vertical gray line ($r=1.18$ au) of the left panels shows the location of the pressure maximum ($\eta = 0$).
({\it bottom row}): $\mathrm{d}^2 \ln P/ \mathrm{d} (\ln r)^2$, which is the left-hand side of eq.~(\ref{eq:pressure_condition}).
The horizontal black dotted line represents 0.
}
\label{fig:12panels}
\end{figure*}

We have shown the five different cases in Figures \ref{aND3} -- \ref{inaDsw3}.
Initial small dust grains successfully grow to planetesimal-size objects in two cases, while in the other three cases most of the solid particles fall onto the central star by the inward radial drift.
In particular, the three MRI inactive cases shows that a small difference of the evolution of $\Sigma_{\rm g}$ sways the fate of the growth of dust grains.
We explore conditions for the dust growth against the radial drift barrier by examining these three cases.

Figure \ref{fig:12panels} presents the radial profiles of various physical quantities of these three MRI inactive cases.
The left (WTb+WM+ZTq) and middle (WTb+WM+STq) panels correspond to the cases with the successful dust growth.
In these cases, the snapshots are taken at the onset time, $t=1.77\times 10^5$ and $1.86\times 10^6$ yrs, respectively, of the growth of dust particles when the Stokes number exceeds St $=1$ at the inner edge of the dust disk, $r = 1.5$ au.
We display snapshots at an arbitrary time, $t=2.5\times 10^5$ yrs, for the right panels (WTb+WM+CTq), which does not yield the significant growth of the solid component.

\subsubsection{Equilibrium Stokes Number}
As we explained earlier (Section \ref{MDWeffect}), the collisional growth initially dominates the radial drift of dust particles, $t_{\rm grow} \ll t_{\rm drift}$, which we call the growth dominated phase.
With the growth of dust grains, $t_{\rm drift}$ ($\propto$ St$^{-1}$ for St $\ll 1$) decreases to eventually give $t_{\rm grow} \sim t_{\rm drift}$.

These features can be seen in the top panels of Figure \ref{fig:12panels}.
$t_{\rm drift}\gg t_{\rm grow}$ in the outer region of $r\gtrsim 50$ au, which is still in the initial growth dominated phase.
However, in the region of $2 \lesssim r \lesssim 50$ au, $t_{\rm drift}$ already approaches to $\approx t_{\rm grow}$. 

We can derive an analytical expression for the Stokes number in the latter phase of the equilibrium state.
Here we consider solid particles in the Epstein regime with $\overline{\alpha_{r\phi}}\ll$ St $\ll 1$.
By equating $t_{\rm drift}$ of eq.~(\ref{eq:t_drift2}) and $t_{\rm grow}$ of eq.~(\ref{eq:t_grow2}), we obtain St in the equilibrium state, 
\begin{equation}
 {\rm St}_{\rm eq} \approx \frac{\sqrt{\pi}}{8}\frac{1}{\eta}\frac{\Sigma_{\rm d}}{\Sigma_{\rm g}}.
\label{eq:St_eq}
\end{equation}

\subsubsection{St $\approx$ St$_{\rm eq}<1$: Radial Drift}
\label{sect:rdrift}
In a typical PPD condition $\eta \sim 10^{-3} - 10^{-2}$ and $\Sigma_{\rm d}/\Sigma_{\rm g} \sim 10^{-3}-10^{-2}$, which give St$_{\rm eq}\sim 0.1$ \citep[][]{Birnstiel2012,Okuzumi2012a,Sato2016a}.
The top panels of Figure~\ref{fig:12panels} also show St $\approx 0.1$ in $2 \lesssim r \lesssim 50$ au. 
(See also St in the bottom panel of Figures \ref{aND3} -- \ref{inaDsw3} at $t=10^{5}$ yrs.)

The dust particles with St$_{\rm eq} < 1$ drift inward before growing to further larger bodies because $t_{\rm drift}\propto$ St$^{-1}$ for St $<1$ and further growth accelerates the radial drift. 
Therefore, these solid particles are in the drift limited state.
Figures \ref{aND3}, \ref{aD3}, and \ref{inaDc3} show that the dust particles of these three cases (STb+NM+ZTq, STb+WM+ZTq, and WTb+WB+CTq) suffer the radial drift and that they do not grow to larger bodies but fall into the central star at $t=10^{6}$ yrs.

\subsubsection{St $\approx$ St$_{\rm eq}>1$: Rapid Growth}
\label{sect:rapidgr}
\begin{figure}
  \centering
  \includegraphics[keepaspectratio, width=8cm, angle=0]{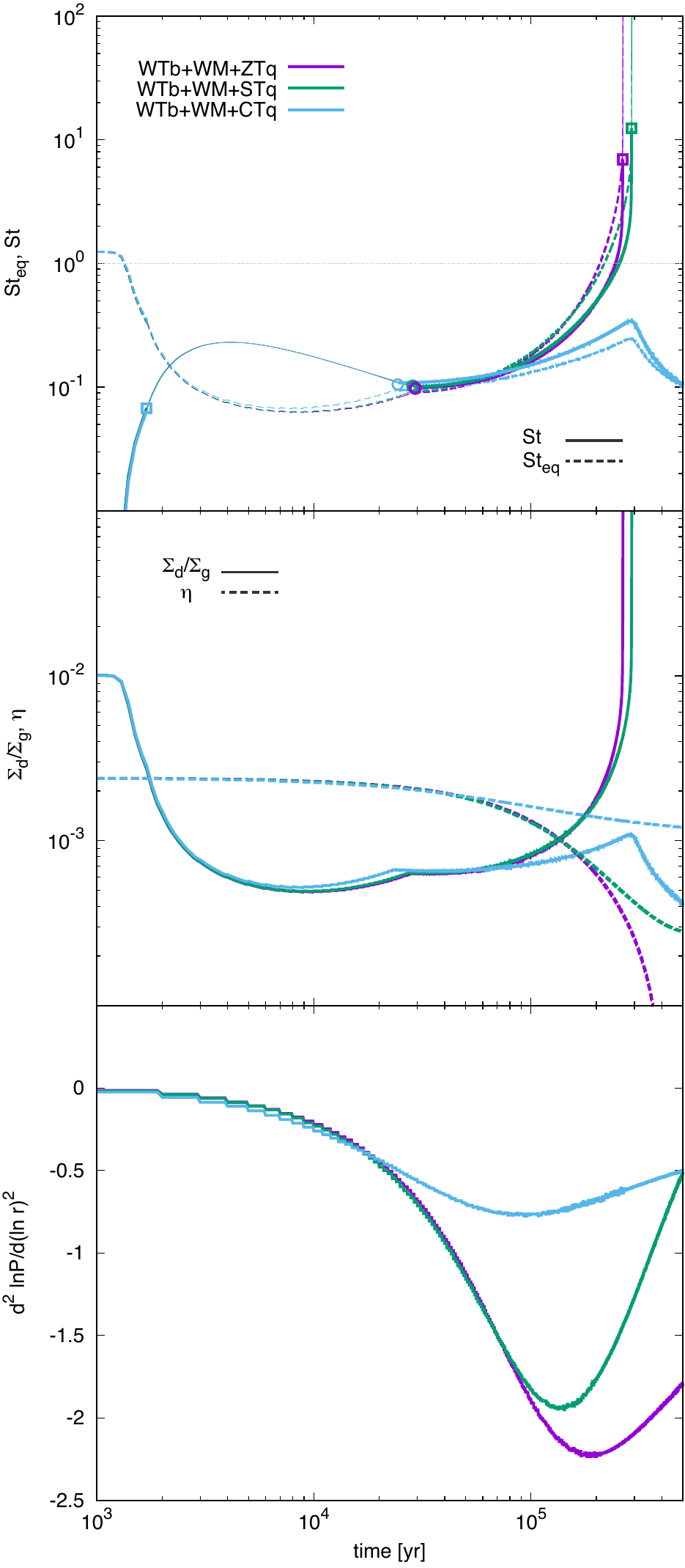}
  \caption{
Time evolution of various physical quantities at $r=1.8$ au for the case with WTb+WM+ZTq (purple), WTb+WM+STq (green), and WTb+WM+CTq (blue).
({\it top}): Stokes number (solid lines) and equilibrium Stokes number (dashed lines).
Squares indicate the transition of the gas drag from the Epstine regime to the Stokes regime; circles correspond to the transition from the Stokes regime to the Epstine regime.
The period during the solid is in the Stokes regime is drawn by narrower lines. 
({\it middle}): Dust-to-gas surface density ratio (solid lines; left axis) and $\eta$ (dashed lines; right axis).
({\it bottom}): $\partial^2\ln P/\partial (\ln r)^2$ (eq.~\ref{eq:pressure_condition}).
}
\label{comp_growth_mech}
\end{figure}

The top left and top middle panels of Figure~\ref{fig:12panels} show that $t_{\rm grow} \ll t_{\rm drift}$ near the inner edge of the solid disk, $r\lesssim 2$ au, which expects the growth of dust particles (Figures \ref{inaD3} and \ref{inaDsw3}).
In order to inspect this growth dominated state we display the time evolution of various physical quantities at $r=1.8$ au of the three MRI inactive cases in
Figure~\ref{comp_growth_mech}.

The top panel of  Figure~\ref{comp_growth_mech} compares the time evolutions of ${\rm St}$ calculated by the coagulation equation (\ref{eq:growth}) and equilibrium ${\rm St}_{\rm eq}$ of eq.~(\ref{eq:St_eq}).
In the initial phase of $t<2\times 10^{3}$ yrs, St ($\ll$ St$_{\rm eq}$) rises up quite rapidly because the dust particles are in the growth dominated state $t_{\rm grow} \ll t_{\rm drift}$ in all the three cases (Note that all the three lines are overlapped).

At $t\approx 2\times 10^{3}$ yrs the state of the gas drag changes from the Epstein regime (thick lines) to the Stokes regime (thin lines) with the increase of particle size, $a$, which is indicated by squares.
Afterward St ($\approx a^2$ in the Stokes regime) gradually decreases until $t<(2-3)\times 10^4$ yrs\footnote{We note that the solid particles are in the drift limited state with the Stokes drag force during $t\approx 2\times 10^3 - 3\times 10^4$ yrs, although St is considerably larger than St$_{\rm eq}$ during most of the period.
The reason why St $>$ St$_{\rm eq}$ is that St$_{\rm eq}$ is derived for the Epstein drag.
}.
This is because smaller particles drift from the outer region by the radial drift.
Therefore, when we watch solid particles at a fixed point of $r=1.8$ au, St is observed to be decreasing with time.

Since the gas density gradually decreases by the MDW and the accretion, the mean free path ($\lambda_{\rm mfp} \propto \rho_{\rm g}^{-1}$ as shown in Section \ref{sect:dust}) increases with time.
As a result, the gas-solid interaction returns back to the Epstein drag at $t\approx (2-3)\times 10^4$ yrs (shown by circles). After that time, the solid particles are in the equilibrium state of St $\approx$ St$_{\rm eq}$.

In all the three cases, St ($\approx$ St$_{\rm eq} \propto \eta^{-1} (\Sigma_{\rm d}/\Sigma_{\rm g})$; eq.~\ref{eq:St_eq}) gradually increases because $\eta$ decreases and $\Sigma_{\rm d}/\Sigma_{\rm g}$ increases (middle panel of Figure~\ref{comp_growth_mech}) , which will be also discussed in Section \ref{sect:condSteq}.  
However,  St ($\approx$ St$_{\rm eq}$) of the purple (WTb+WM+ZTq) and green (WTb+WM+STq) lines start to deviate upward from the light blue (WTb+WM+CTq) line.
In these two cases St finally jumps up beyond unity. 

For St $>1$, the radial velocity of solid particles follows $v_{{\rm d},r}\propto$ St$^{-1}$ (see eq.~\ref{eq:vdr} and Section \ref{sect:rdtime}). 
Therefore, the growth of solid particles with ${\rm St}>1$ slows down the inward radial drift. 
If a region with St $>1$ is formed locally, it accumulates solid particles that drift inward from the outer part of a PPD.
This enhances the dust density there.
The local enhancement of $\Sigma_{\rm d}$ causes the faster growth of solid particles (eq.~\ref{eq:t_grow1}).
In summary, an increase of St finally leads to a further increase of St once St exceeds unity; a positive feedback sets in to trigger the growth of solid particles.

The purple and green lines in the upper and middle panels of Figure~\ref{comp_growth_mech} clearly show this behavior.
After St exceeds unity, $\Sigma_{\rm d}/\Sigma_{\rm g}$ jumps up irrespective of the evolution of $\eta$.
The accumulation of the solid forms the ring-like structures seen in Figures \ref{inaD3} and \ref{inaDsw3}. 

\subsubsection{Requirements for St $\approx$ St$_{\rm eq} > 1$}
\label{sect:condSteq}
We have shown that the key for drifting particles to reach the growth dominated state is to achieve St$_{\rm eq}\gtrsim 1$ in the equilibrium state.
Since ${\rm St}_{\rm eq} \propto \eta^{-1} (\Sigma_{\rm d}/\Sigma_{\rm g})$ (eq.~\ref{eq:St_eq}), smaller $\eta$ and larger $\Sigma_{\rm d}/\Sigma_{\rm g}$ favor larger St$_{\rm eq}$.

The middle panel of Figure~\ref{comp_growth_mech} shows that $\eta$ decreases with time because the MDW disperses the gas in an inside-out manner to reduce the outward pressure gradient (Figures \ref{inaD3}--\ref{inaDsw3}).
In particular, the two successful cases (purple and green lines) show a more rapid drop of $\eta$ to give larger St$_{\rm eq}$ than the unsuccessful case (light blue line).
The consistent trend is obtained for the snapshot radial profile of $\eta$ in Figure~\ref{fig:12panels}; the third left (WTb+WM+ZTq) and third middle (WTb+WM+STq) panels show that $\eta$ deviates downward from the initial profile $\propto r^{1/2}$ and gives quite small $\eta < 10^{-3}$ in $r\lesssim 2$ au.
The left case even exhibits $\eta<0$ inside $r = 1.18$ au because of the pressure bump. 

Although the dust-to-gas ratio, $\Sigma_{\rm d}/\Sigma_{\rm g}$, initially drops from the initial value of $10^{-2}$ to $\lesssim 10^{-3}$ by the radial drift, $\Sigma_{\rm d}/\Sigma_{\rm g}$ slowly rises after $t\gtrsim 3\times 10^4$ yrs (middle panel of Figure~\ref{comp_growth_mech}).
\replaced{$\Sigma_{\rm d}/\Sigma_{\rm d}$}{$\Sigma_{\rm d}/\Sigma_{\rm g}$} of the two successful cases (purple and green lines) increases more rapidly to $>10^{-3}$, which also contributes to reaching St ($\approx$ St$_{\rm eq}$) $> 1$. 

From these arguments, we can obtain a condition to trigger the subsequent growth of solid particles in the drift limited state with St $\gtrsim 0.1$,
\begin{equation}
  \Sigma_{\rm d}/\Sigma_{\rm  g} \gtrsim \eta. 
  \label{eq:condrg}
\end{equation}
The middle panel of Figure~\ref{comp_growth_mech} demonstrates that the solid line ($\Sigma_{\rm d}/\Sigma_{\rm  g}$) of the two successful cases (purple and green) overtakes the corresponding dashed line ($\eta$), while this does not occur in the unsuccessful case (light blue).
We can conclude that eq.~(\ref{eq:condrg}) is a reasonable criterion to reach the growth dominated phase.
Although eq.~(\ref{eq:condrg}) is a rearranged expression of St$_{\rm eq} \gtrsim 1$ with eq.~(\ref{eq:St_eq}) apart from the numerical factor, eq.~(\ref{eq:condrg}) is important as an independent condition because eq.~(\ref{eq:St_eq}) is valid only for St$\ll 1$ in a strict sense. 
Our numerical calculations have confirmed that $\frac{1}{\eta}\frac{\Sigma_{\rm d}}{\Sigma_{\rm g}}$ is a useful indicator even for relatively large St$\sim 1$. 

After St ($\approx$ St$_{\rm eq}$) of WTb+WM+CTq (light blue) slowly increases up to $\approx 0.35$ at $t\approx 3\times 10^5$ yrs (top panel of Figure~\ref{comp_growth_mech}), St decreases with $\Sigma_{\rm d}/\Sigma_{\rm g}$ after that time (middle panel).
This is because the dust grains that are initially located at the outer boundary of $r=200$ au already reach $r=1.8$ au at this time by the radial drift and the mass supply from the outer region ceases afterward.
If the initial radius of the dust disk was larger than the current setup, this case could also reach the growth dominated state of solid particles at a later time.

We have explained that both the decreasing $\eta$ and increasing $\Sigma_{\rm d}/\Sigma_{\rm g}$ play important roles in obtaining ${\rm St}_{\rm eq} \gtrsim 1$ in the two cases with the significant dust growth.
The small $\eta$ is a characteristic consequence of the inside-out evacuation of the gas by the MDW.
On the other hand, it is not straightforward to understand the increase of $\Sigma_{\rm d}/\Sigma_{\rm g}$.
While we discussed that the positive feedback loop inevitably enhances $\Sigma_{\rm d}/\Sigma_{\rm g}$ for St $>1$, the requirement here is to accumulate solid particles with St $<1$, which we consider below.

\subsubsection{Dust Accumulation under St $<1$}
\label{sec:cond-dust-surf}

Let us introduce the radial mass flux of dust particles:
\begin{eqnarray}
\label{eq:Md}
\dot{M}_{\rm d}=2\pi r \Sigma_{\rm d} v_{{\rm d},r},
\end{eqnarray}
where $\dot{M}_{\rm d}<0$ when dust particles drift inward.
When $\partial \dot{M}_{\rm d}/\partial r < 0$ at a single annulus in a disk, the net mass flux into this annulus is positive so that $\Sigma_{\rm d}$ increases there.
We derive an analytic expression that gives this condition for the local pile-up of dust particles below.

Let us consider dust particles in the drift limited state with St $\approx 0.1$ (Figure~\ref{fig:12panels} and Section \ref{sect:rdrift}).
When St is a constant, $v_{{\rm d},r}$ is controlled only by the the gas pressure profile so that we have (see eq.~(\ref{eq:F2}) of Appendix~\ref{sec:deriv-cond-dust})
\begin{eqnarray}
\label{eq:Md_dep}
\dot{M}_{\rm d} \propto (r \Sigma_{\rm d})\frac{\partial \ln P}{\partial \ln r},
\end{eqnarray}
where we here used the radial dependence of the gas temperature, $T \propto r^{-1/2}$, of the MMSN.

If we adopt a power-law dependence of $\Sigma_{\rm d} \propto r^{-q_{\rm d}}$ in eq.~(\ref{eq:Md}), the condition for the accumulation, $\partial \dot{M}_{\rm d}/{\partial r}<0$, reads (see Appendix~\ref{sec:deriv-cond-dust} for the derivation),
\begin{eqnarray}
\label{eq:pressure_condition}
\frac{\partial^2\ln P}{\partial(\ln r)^2} < (q_{\rm d}-1)  \frac{\partial \ln P}{\partial \ln r}.
\end{eqnarray}
In the drift limited phase, the radial dependence of $\Sigma_{\rm d}$ tends to approach to the steady state value, $q_{\rm d}=1$ \citep[][see also Appendix~\ref{sec:deriv-cond-dust}]{Birnstiel2012}, which gives a practical criterion,
\begin{eqnarray}
\label{eq:press_condition_simple}
\frac{\partial^2\ln P}{\partial(\ln r)^2} < 0,
\end{eqnarray}
to create a local concentration of dust particles;
the convex upward profile of $P$ drives converging dust flows. 

We show the radial profile and the time evolution of $\frac{\partial^2\ln P}{\partial (\ln r)^2}$ in the bottom panels of Figure~\ref{fig:12panels} and the bottom panel of Figure~\ref{comp_growth_mech}, respectively.
These figures illustrate that the WTb+WM+NTq and +STq cases give small $\frac{\partial^2\ln P}{\partial(\ln r)^2}\approx -2$ at $r\sim 1.5-2$ au, which can induce strong convergent dust flux.
On the other hand, $\frac{\partial^2\ln P}{\partial(\ln r)^2} (> -1)$ of the WTb+WM+CTq case is not so small as those of the two cases, and therefore, the excited convergent flows are too weak to proceed subsequent growth of solid particles.

An important point is that a pressure bump structure, which corresponds to the position of $\partial P/\partial r=0$, is not necessarily required to excite strong convergent dust flux, which is driven by negative $\partial^2 P/\partial r^2$.
Figure~\ref{fig:12panels} shows that $\eta > 0$ in the entire region of the WTb+WM+STq case (middle column), namely there is no local maximum of the gas pressure.
This case clearly demonstrates that dust particles can be piled up by the strong converging flux even without any pressure bump.
The WTb+WM+ZTq case (left column of Figure~\ref{fig:12panels}) also illustrates that the location ($r\approx 1.5$ au) of the solid concentration does not coincide with the pressure bump at $r=1.18$ au.

\begin{figure}
  \centering
  \includegraphics[keepaspectratio, width=8cm, angle=0]{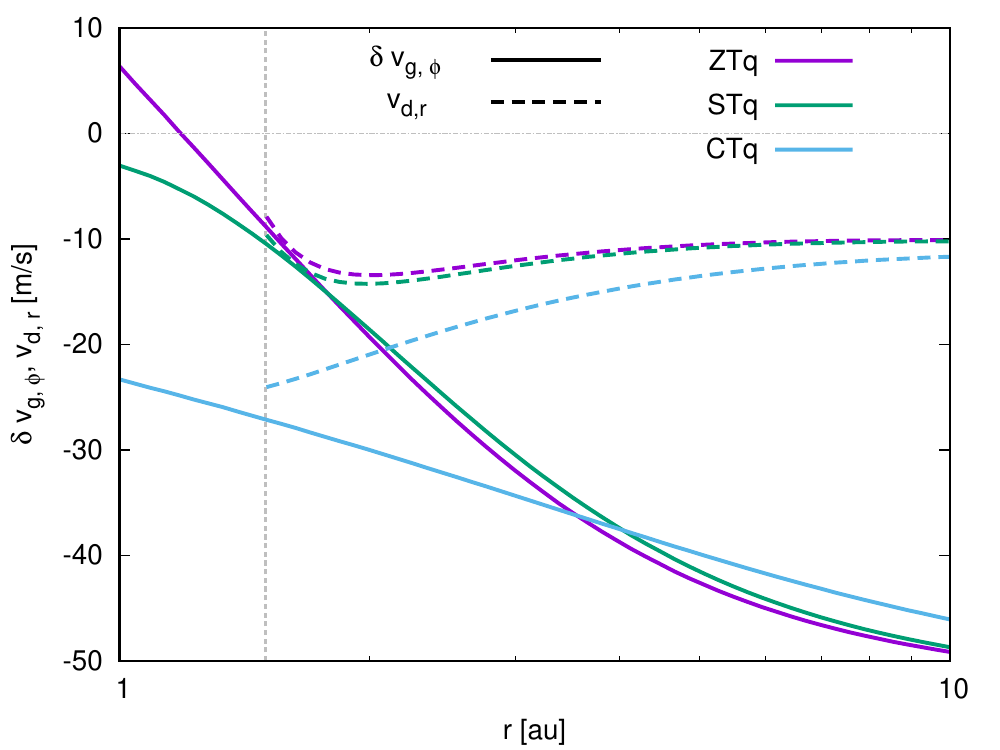}
 \caption{
 \explain{This is a new figure.}
 \added{
Radial profile of the dust radial velocity (dashed lines) and deviation of gas azimuthal velocity from the Keplerian velocity (solid lines) for the case with WTb+WM+ZTq (purple), WTb+WM+STq (green), and WTb+WM+CTq (blue).
Each case shows a snapshot of the same time as shown in Figure~\ref{fig:12panels}.
The vertical gray dashed line depicts the inner boundary for the dust component, $r=1.5$ au.
 }
}
\label{fig:vphi}
\end{figure}

\added{
We compare the deviation of the azimuthal velocity of the gas component from the Keplerian one, $\delta v_{{\rm g},\phi} = v_{{\rm g},\phi} - v_{\rm K} = \eta v_{\rm K}$, (solid lines) and the radial velocity of the solid component, $v_{{\rm d},r}$, (dashed lines; eq.~\ref{eq:vdr}) in Figure \ref{fig:vphi}, which further supports our claim.
Since both $\delta v_{{\rm g},\phi}$ and $v_{{\rm d}, r}$ are negative in $r \ge 1.5$ au, there is no pressure maximum so that dust particles drift inward.

The two successfull cases (purple and green lines in Figure~\ref{fig:vphi}) give $v_{{\rm d},r} \sim \delta v_{{\rm g},\phi}$ in $r \lesssim 2$ au.
From eq.~(\ref{eq:vdr}) we obtain $|v_{{\rm d}, r}| \leq |\delta v_{{\rm g}, \phi}|$ for $v_{{\rm g}, r} \ll v_{{\rm d}, r}$.
Therefore $v_{{\rm d}, r}$ almost reaches its maximum value that is given for ${\rm St} \sim 1$ in $r \lesssim 2$ au.
$v_{{\rm d},r}$ in this region shows converging dust flux, which leads to the pile-up of $\Sigma_{\rm d}$, whereas we recall that the condition for the pile-up is $\mathrm{d}\dot{M}_{\rm d}/\mathrm{d}r<0$ (eq.~\ref{eq:Md}) in a strict sense.
Since $v_{{\rm d},r}$ follows $\delta v_{{\rm g},\phi}$ that is determined by the profile of $\Sigma_{\rm g}$ there, the pile-up of $\Sigma_{\rm d}$ is triggered by the evolution of the MDW-governed PPDs.

As a result of the pile-up of $\Sigma_{\rm d}$, the condition for the rapid dust growth (eq.~\ref{eq:condrg}) is easily achieved in this region where ${\rm St} \sim 1$.
The radial drift barrier is overcome even without the presence of a pressure maximum.
The difference between the CTq case (blue lines in Figure~\ref{fig:vphi}) and the successful two cases is the deviation of $\Sigma_{\rm g}$ from the initial power-law profile by MDWs.
For the CTq case, $v_{{\rm d},r} \sim \delta v_{{\rm g},\phi}$ is not achieved because the radial profile of $\Sigma_{\rm g}$ is not so deviated from the initial profile as that of the other two cases.
}

\begin{figure*}
   \centering
  \includegraphics[keepaspectratio, width=18cm, angle=0]{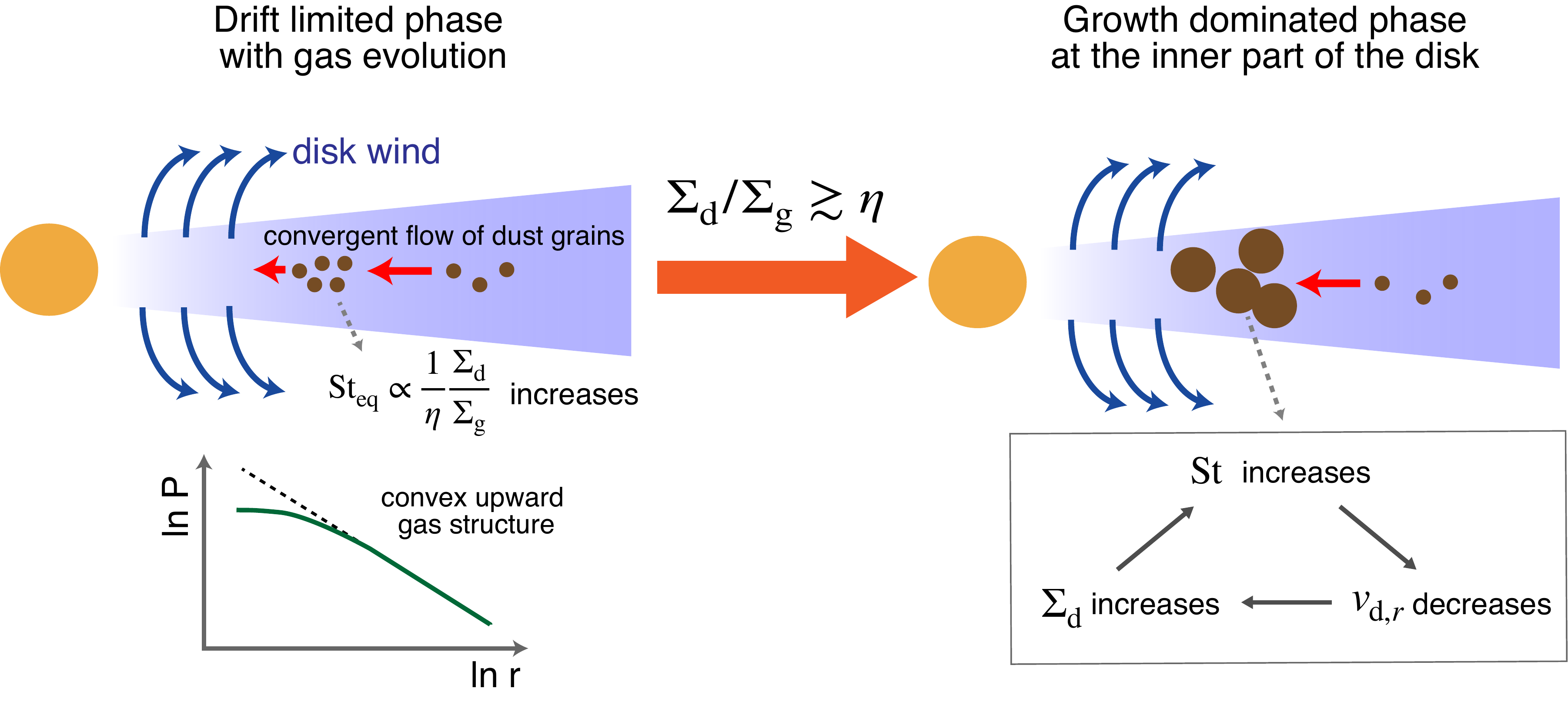}
  \caption{
    Schematic illustration of our new growth mechanism of dust particles.
    ({\it left}): Solid particles are in the drift limited state with St $\sim 0.1$. 
    They are eventually accumulated at the location where $P$ has a convex upward radial profile.
    Simultaneously St $\approx {\rm St}_{\rm eq}$ gradually increases by the increase of $\Sigma_{\rm d}/\Sigma_{\rm g}$ and the decrease of  $\eta$.
    ({\it right}): Solid particles in the inner part of a PPD enter the growth dominated phase.
    When $\Sigma_{\rm d}/\Sigma_{\rm g} \gtrsim \eta$, i.e., St $\approx {\rm St}_{\rm eq} \gtrsim 1$, a positive feedback loop among ${\rm St}$, $\Sigma_{\rm d}$, and $v_{{\rm d}, r}$ triggers the subsequent growth of dust particles.
}
\label{fig:schematic}
\end{figure*}

\subsubsection{Summary of Dust Growth}
\label{sec:summ-growth-proc}

We summarize the overview of the mechanism for the growth of solid particles in Figure~\ref{fig:schematic}.
The dust growth consists of two steps:
(1) Converging dust flux gradually forms a local concentration of solid particles by the supply of radially drifted dust grains from the outer region with the increase of their St $\approx {\rm St}_{\rm eq}\propto \frac{1}{\eta}\frac{\Sigma_{\rm d}}{\Sigma_{\rm g}}$ (Sections \ref{sect:condSteq} and \ref{sec:cond-dust-surf}): 
(2) Once St ($\approx$ St$_{\rm eq}$) $\gtrsim 1$, the unstable growth mode of solid particles sets in by the positive feedback (Section \ref{sect:rapidgr})
\added{\footnote{\added{We here use ``unstable'' because of the clear evidence of the positive feedback loop among $v_{{\rm d},r}$, $\Sigma_{\rm d}$, and St, although we have not carried out the linear perturbation analysis for this problem. }}}.

We point out that the step (1) could be accomplished by different processes from the MDW, provided that they create a region with small $\eta$ and large $\Sigma_{\rm d}/\Sigma_{\rm g}$ in a drift limited region.
We discuss other possibilities that can trigger unstable dust growth in Section~\ref{sec:rapid-dust-growth}.

We would like to claim that our new growth mechanism is different from those proposed previously.
In our process, all the dust grains firstly experience the drift limited state.
After that, part of them switch to the growth dominated state in the Epstein regime.
These features are in contrast to the mechanism introduced by \citet{Okuzumi2012a}, in which fluffy dust grains skip the drift limited phase and directly enter the growth dominated phase in the Stokes regime.
Our mechanism is also different from that driven by the backreaction from dust to gas that decelerates the radial drift of dust grains  \citep{2016A&A...594A.105D, 2017MNRAS.467.1984G}.

\subsection{Total mass of solid component}
\label{massloss}
We inspect how much fraction of the initial solid mass ($103M_{\rm E}$ between 1.5 au and 200 au) survives in the calculated disks.
The dust grains are dispersed via (1) the radial drift to the central star and (2) the upward entrainment by MDWs.
We found that in all the eight cases the former dominates the latter, because in the inner region where the effect of the MDW is most prominent the dust grains grow so rapidly that they are too large to be dragged upward.

\begin{figure}
\begin{center}
 \includegraphics[width=8cm, clip]{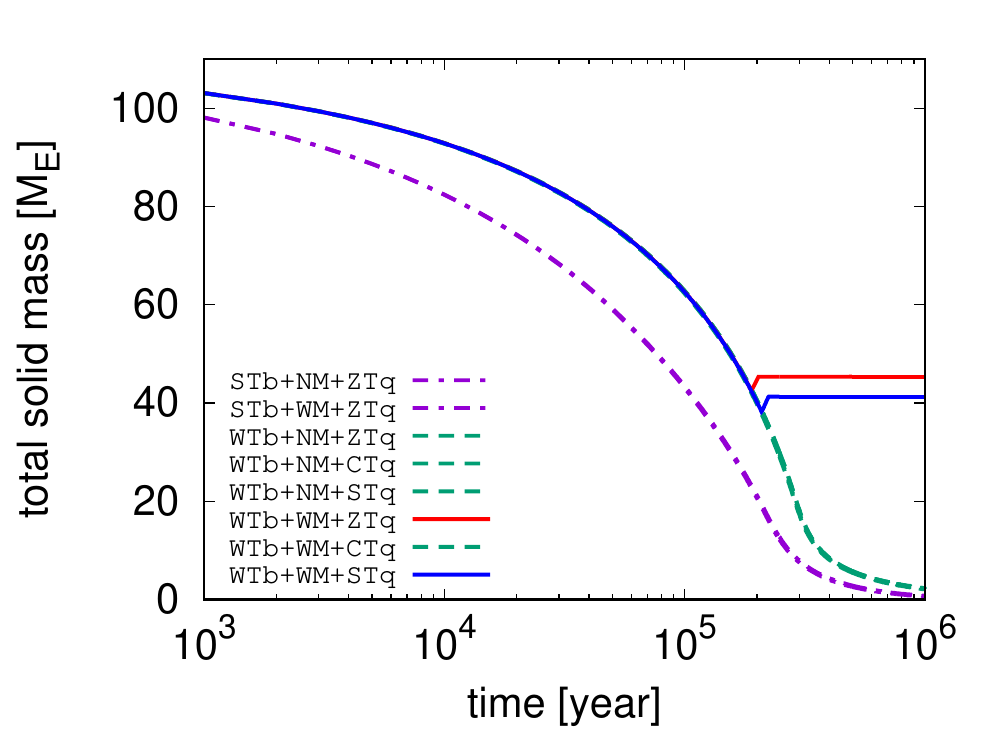}
  \caption{
The time evolution of the total mass normalized by $M_{\rm E}$ of the solid component in the computational domain.
The two MRI active cases (STb+XM+ZTq) are plotted together by a purple dot-dashed line because the difference between the two cases is less than 1\%.
The four MRI inactive cases (WTb+NM+XTq \& WTb+WM+CTq) that do not show the subsequent growth of solid particles are also plotted together by a green dotted line.
The red solid and blue solid lines denote the cases of WTb+WM+Ztq and WTb+WM+STq, respectively. 
}
\label{fig:massloss}
\end{center}
\end{figure}

Figure~\ref{fig:massloss} presents the time evolution of the total solid mass left in the computational domain of all the cases tabulated in Table \ref{parameter}.
The time evolutions can be classified into three types: (i) MRI-inactive cases that show significant dust growth (solid lines), (ii) MRI-inactive cases without significant dust growth (dashed lines), and (iii) MRI-active cases without significant dust growth (dot-dashed lines). 
In type (i), when the subsequent dust growth sets in at the inner edge of the solid disk, there is $\sim 4\%$ numerical error, which appears as a spike at $t\approx 2\times 10^5$ yrs. 

Figure~\ref{fig:massloss} shows that in types (ii) and (iii) most of the initial solid mass is lost at $t=10^6$ yrs because of the radial drift.
On the other hand, the two cases of type (i) leaves the solid mass of $\sim 40$ -- $45 M_{\rm E}$ in the disk.
The remained mass is determined by the time when St exceeds $\approx 1$ in the local dust concentration formed near the inner edge.
This is because, while before that time all the solid particles drift inward to the central star, after that time they are captured by the ring-like concentration of the solid component. 
As shown in Figure~\ref{comp_growth_mech}, St reaches $\approx 1$ at slightly earlier time in the WTb+WM+ZTq case (red solid line) than in the WTb+WM+STq case (blue solid line).
Therefore, the former case leaves the slightly larger solid mass. 

The remained mass of $40-45 M_{\rm E}$ is sufficient to 
bear the total solid component in the planets of the solar system and probably a sizable fraction of exoplanetary systems.
The sufficient gaseous mass is also considered to remain to form gas giants when the subsequent growth of dust grains occurs at $t\approx 2\times 10^5$ yrs.

\section{Discussion}
\label{discussion}
\added{We discuss the universality and applicability of the conditions for the dust growth presented so far in Section~\ref{sec:univ-new-growth}.}
We assumed several approximations when we solved the coagulation equation of solid particles and the evolution of $\Sigma_{\rm g}$ and $\Sigma_{\rm d}$.
We discuss limitations of our treatment and processes that are not considered in Sections~\ref{sec:gas-component}-~\ref{sec:rapid-dust-growth}.
We also discuss observational implications in Section~\ref{sec:impl-some-observ}.

\added{
\subsection{Universality of the new growth mechanism}
\label{sec:univ-new-growth}

We introduced the condition for the pile-up of dust particles (eqs. \ref{eq:pressure_condition} \& \ref{eq:press_condition_simple}) and the condition for the rapid dust growth (eq. \ref{eq:condrg}) in the previous section.
Although in this paper we have focused on the evolution of PPDs with MDWs, these conditions are also applicable to general PPDs that are subject to other physical processes.

The pile-up condition (eqs. \ref{eq:pressure_condition} \& \ref{eq:press_condition_simple}) is satisfied by various physical processes, non-ideal MHD effects (Section~\ref{sec:gas-component}), photoevaporatibe winds (PEWs; Section~\ref{sec:phot-wind}), Rossby wave instability (RWI; Section~\ref{sec:multi-dimens-effects}), and pressure bumps (Section~\ref{sec:rapid-dust-growth}).
ln order that the rapid growth condition (eq. \ref{eq:condrg}) is satisfied during the drift-limited state, the pile-up condition should be maintained for a sufficiently long time to accumulate solid particles that drift inward from the outer region.
If the later disk evolution achieves the pile-up condition, most of the  solid particles have already infallen to the central star, resulting in the insufficient pile-up to satisfy the rapid growth condition.

In our analysis we have assumed the axisymmetric approximation.
Therefore, we have to examine how non-axisymmetric modes affect our axisymmetric growth mode.
On one hand non-axismmetric instability, e.g. RWI (Section \ref{sec:multi-dimens-effects}), may further support our growth mode.
On the other hand azimuthal disturbances may saturate the nonlinear growth of our process, which will be studied in our future work.
}

\subsection{Uncertainty of MHD effects}
\label{sec:gas-component}
\explain{We changed the title of this subsection from ``Uncertainty of Parameters''.}
We solved the evolution of the gas component by the simple model of eq.~(\ref{eq:gas}) with the three parameters, $\overline{\alpha_{r\phi}}$, $\overline{\alpha_{\phi z}}$, and $C_{\rm w}$.

$\overline{\alpha_{r\phi}}$ is mainly determined by the ionization degree \citep{2000ApJ...543..486S,2011ApJ...732L..30H,2016ApJ...817...52M}.
Although we assumed a constant $\overline{\alpha_{r\phi}}$, in realistic situations $\overline{\alpha_{r\phi}}$ varies with $r$ because an MRI-inactive dead zone is formed \citep{1996ApJ...457..355G,1999ApJ...515..776S}.
In this case pressure bumps could be created at the edges of the dead zone \citep{2008A&A...491L..41L,Suzuki2010a,2010A&A...515A..70D} and dust grains could grow into larger bodies there.
The wind torque, $\overline{\alpha_{\phi z}}$, may also have a large uncertainty concerning non-ideal MHD effects \citep{Bai2017a}.

The mass loss, $C_{\rm w}$, by the MDW also contains uncertainties.
\citet{Suzuki2016a} argued that $C_{\rm w}$ could be constrained by the released gravitational energy through the accretion, which reduces $C_{\rm w}$ in the inner region.
However, this constraint can be loosened if the energy is supplied by external heating owing to the irradiation from the central star.
Therefore, we adopted the simplest situation of constant $C_{\rm w}$ in the present paper.

\added{
Various non-ideal MHD effects potentially form substructures in PPDs.
The ambipolar diffusion develops zonal flows that induce local pressure maxima in the relatively outer part of PPDs \citep{2017A&A...600A..75B,Suriano2018a,Suriano2019}.
The Hall effect also develops zonal flows and leads to the formation of large scale vortices in MRI active zones\citep{2016A&A...589A..87B,2018ApJ...865..105K}.

These substructures are formed on short timescales of a few to dozens of local orbital times and prevent the radial drift of dust grains.
As is expected, they are suitable sites for the growth of dust grains; we can apply the same analysis based on the pileup condition (eqs. \ref{eq:pressure_condition} \& \ref{eq:press_condition_simple}) and the rapid growth condition (eq. (\ref{eq:condrg})) to these substructures.
}

\added{
\subsection{Photoevaporative wind}
\label{sec:phot-wind}

Photoevaporative winds (PEWs) are also believed to be a promising mechanism in the dispersal of PPDs.
The gas of PPDs is heated up by high-energy (X-ray, EUV, FUV; XEFUV) photons, and eventually escape from the surface through thermal driven outflows \citep[e.g.,][]{Alexander2006a,2010MNRAS.401.1415O,2009ApJ...705.1237G}.
The mass loss rate by PEWs is regarded to be spatially dependent so that gap/hole structures are possibly formed in PPDs, whereas the detailed mass loss profile is still under debate \citep{2001MNRAS.328..485C}.

\citet{2020MNRAS.492.3849K} calculated evolution of the gas of PPDs including both MDWs and XEUV PEWs.
They showed that the evolution of PPDs is primarily controled by MDWs at early times $\lesssim 10^6$ yrs, while the contribution from PEWs is gradually significant after $> 10^6$yrs.
Our results show that the dust growth occurs within a few $10^5$yrs (Section \ref{results}), during which the global evolution of the PPDs is subject to MDWs rather than PEWs.

\citet{2015ApJ...804...29G} calculated the evolution of PPDs with XEFUV PEWs with explicitly taking into account the evolution of the grain size.
Since they did not consider MDWs, they adopted relatively strong turbulence, $\alpha \sim 10^{-3} - 10^{-2}$, in order to reproduce the typical lifetime of PPDs.
Therefore the dust component in their model is in the fragmentation-limited state, in which the maximum size of dust grains is controlled by collisional fragmentation, because of the large relative velocity between dust particles (see also Sec.~\ref{sec:fragmentation}).
In contrast, when MDWs are active, a small $\alpha \sim 10^{-4}$ is sufficient to explain the typical lifetime.
Hence, the collision velocity is slower than that used in \citet{2015ApJ...804...29G}, and the dust component is in the drift limited state in our calculations (Section \ref{results}).

The growth of dust grains also affects the properties of PEWs \citep[e.g.,][]{2015ApJ...804...29G,2018ApJ...857...57N,2018ApJ...865...75N}.
Therefore, it is important to study mutual relations among PEWs, MDWs, and dust evolution in future work.
}

\subsection{Backreaction on Gas}
\label{sec:backreaction-gas}
We solved the main equations neglecting the backreaction from dust to gas.
However, the backreaction is important when the dust-to-gas ratio is high $\gtrsim 0.1$ \citep{Nakagawa1986a}.
If the backreaction is included, the radial velocity of solid particles is modified as
\begin{eqnarray}
 v_{{\rm d}, r} \approx \frac{-2{\rm St}}{(1+\rho_{\rm d}/\rho_{\rm g})^2 + {\rm St}^2}\eta v_{\rm K}, 
\end{eqnarray}
where $\rho_{\rm d}$ is volumetric density of dust grains.
This equation indicates that the drift speed is reduced for a large dust-to-gas ratio irrespective of the direction of $v_{{\rm d}, r}$.
Therefore, once dust grains are accumulated, the radial drift is suppressed, which further supports the growth of dust particles.

The suppression of $v_{{\rm d},r}$ is more severe for a higher dust-to-gas ratio. 
Figures~\ref{inaD3} and \ref{inaDsw3} show the ring with high $\Sigma_{\rm d}/\Sigma_{\rm g} > 1$.
\deleted{
In such a situation, the outward migration of large solid particles, which is seen in Figure~\ref{inaD3}, will not occur in realistic situations with the backreaction.
Instead, the ring-like structure would stay at the same place for a rather long time.
}
\added{
As described in Section~\ref{dustring}, the radial drift of dust grains in the ring-like regions is already quite slow with timescale of $\sim 10^{9}$ yr, which is sufficiently longer than the disk lifetime ($\sim 10^{6-7}$ yrs).
The backreaction further slows down the radial drift of the solid particles that make up the ring-like regions.
}

The backreaction also affects the profile of gas density.
\citet{Taki2016a} found from their local simulations that the dust accumulation at a pressure bump flattens the radial profile of the gas by the redistribution of the angular momentum between the dust and the gas through the backreaction.
The profile of $\Sigma_{\rm g}$ would be also flattened by the backreaction on the gas in the WTb+WM+ZTq and WTb+WM+STq cases (Figures~\ref{inaD3} and \ref{inaDsw3}).
However, this does not severely affect the onset of the subsequent growth of dust particles because it is triggered for small $\Sigma_{\rm d}/\Sigma_{\rm g} < 10^{-2}$ (middle panel of Figure~\ref{comp_growth_mech}).
It is worth estimating the dust-to-gas mass ratio, $\rho_{\rm d}/\rho_{\rm g}$, because it is more directly related to dust-gas interactions.
$\rho_{\rm d}/\rho_{\rm g}$ at the midplene can be derived from eqs.~(\ref{eq:rhomidplane}) and (\ref{eq:hd}) as
\begin{eqnarray}
 \frac{\rho_{\rm d}}{\rho_{\rm g}} = \left(\frac{\Sigma_{\rm d}}{\Sigma_{\rm g}}\right)\left(1+\frac{\rm St}{\overline{\alpha_{r\phi}}} \frac{1+2{\rm St}}{1+{\rm St}} \right)^{1/2},
\end{eqnarray}
where we used $\rho_{\rm d} = \Sigma_{\rm d}/(\sqrt{2\pi}h_{\rm d})$.
We employ the typical values, $\Sigma_{\rm d}/\Sigma_{\rm g} \approx 10^{-3}$, $\overline{\alpha_{r\phi}} = 8\times 10^{-5}$, and ${\rm St}\approx 1$, which are taken from the onset time of the dust growth (Figure~\ref{comp_growth_mech}), and then we obtain $\rho_{\rm d}/\rho_{\rm g} \sim 0.1$ at the midplane.

The modification of the gas density profile again suppresses the saturation level of dust density for $\rho_{\rm d}/\rho_{\rm g}\gtrsim 1$, which is also confirmed in a 2D global simulation \citep{Kanagawa2018a}.
Although a spike of $\Sigma_{\rm d}$ is seen at the outer edge of the growth dominated region in the middle panel of Figures~\ref{inaD3} and \ref{inaDsw3}, it would be smoothed out in realistic situations with the backreaction on the gas.

\added{
The gas velocity is also affected by the backreaction.
When dust particles drift inward, the angular momentum of gas increases.
The gas component moves outward near the midplane.

The outward gas flow affects the dust evolution for smaller grains which are tightly coupled to the gas \citep{2018MNRAS.479.4187D,2019ApJS..241...25B}.
In our model, especially in the ring-like region, such dust grains tightly coupled to the gas are a minor population of dust grains because the growth of dust grains dominates the dust evolution (see also Appendix~\ref{app:single_size_approximation}).
Therefore only a small number of dust grains can be delivered to the outside of the ring-like region.
These dust grains delivered by the outward gas flow may contribute to some observational features.
}

\subsection{Fragmentation}
\label{sec:fragmentation}
Although we do not consider the effect of collisional fragmentation in Equation~(\ref{eq:growth}), it is effective for high speed collisions.
Collisional coagulation of dust grains generates dust aggregates.
The critical collisional fragmentation velocities $v_{\rm cr}$ for dust aggregates are investigated with $N$-body simulations \citep{Wada2013a}.
The critical velocities are estimated to be 60-80\, m/s for icy dust and 6-8\,m/s for silicate dust, whereas $v_{\rm cr}$ is still very uncertain and it is reported that $v_{\rm cr}$ could be as fast as $50 \ {\rm m}/{\rm s}$ \citep{Kimura15, Steinpilz19}.
Here we estimate the collisional velocity in our calculations to be compared with the critical velocities.

During dust growth, the relative velocities have maximum values at ${\rm St} \approx 1$.
Successful dust growth takes place in the region with small $\eta$.
Turbulence mainly determines the relative velocities because the radial drift is negligible there.
The maximum velocities arising from the turbulence are estimated to be $\sim \sqrt{\overline{\alpha}} c_{\rm s}$ from eq.~(\ref{eq:vI}) with ${\rm St}_1 \approx 1$.
For the cases with successful dust growth, where the turbulence is weak, the maximum turbulent
velocity is $\sim 9 (r/{\rm 1\,au})^{-1/4}$\,m/s, which is much slower than the critical velocity for icy aggregates.

In the present calculations we fix $\Sigma_{\rm d}=0$ inside the snowline to avoid the treatment of silicate dust (Section~\ref{numerical}).
However, if we take into account collisional fragmentation properly,
\added{
the maximum size of silicate dust grains is limited by the fragmentation inside the snowline.
In this case, the unrealistic growth of dust grains caused by the initial condition does not occur at the beginning of calculations.
Thus }
we may handle the evolution of the solid component in $r<1.5$ au.

\subsection{Hydrodynamical Instabilities}
\label{sec:multi-dimens-effects}

In this paper, we do not solve hydrodynamical equations directly.
PPDs that evolve with MDWs, however, potentially drive hydrodynamical instabilities by the modification of the profile of density and gas pressure.

It is well known that an axisymmetric rotational instability, which is called ``Rayleigh's criterion'', occurs when $\kappa^{2}(r)<0$, where $\kappa (r)$ is the epicyclic frequency \citep{1960PNAS...46..253C}.
When the radial profile of the gas pressure has a \replaced{narrow (strong) pressure bump}{sharp pressure bump}, its inner side may break this condition.

In addition, when the radial profile of vortensity has a local minimum, Rossby-wave instability (RWI), which is a type of non-axisymmetric hydrodynamical instabilities in differentially rotating disks, sets in \citep{1999ApJ...513..805L,2016ApJ...823...84O,2018ApJ...864...70O}.
The instability criterion can be fulfilled even for a shallower slope of the gas pressure than that for the Rayleigh's criterion.
Therefore, the MDW possibly favors the RWI in PPDs.
The RWI forms large anticyclonic vortices, which collects dust particles because it is a non-axisymmetric pressure bump \citep{2009A&A...497..869L}.
\added{
Our proposed mechanism for grain growth is not exclusive of the mechanism of RWI vortices; they possibly operate in an cooperative manner.
}

\subsection{Pressure Bump}
\label{sec:rapid-dust-growth}

Various potential mechanisms have been proposed to form a pressure bump in PPDs because it is favorable sites for the formation of planetesimals \citep[e.g.,][]{2014prpl.conf..547J}.
Inhomogeneous mass accretion rate with radial distance possibly creates a pressure bump at the inner edge of an MRI dead-zone \citep{2008A&A...491L..41L} and at the H$_2$O snow line \citep{2007ApJ...664L..55K} (see also Section~\ref{sec:introduction}).

We demonstrated that a pressure bump is not necessarily required but a convex upward profile, $\partial^2\ln P/\partial (\ln r)^2 < 0$, can pile up dust particles to trigger the subsequent growth to larger solid bodies.
In other words, our work loosens the condition for the planetesimal formation.

\subsection{Implication for Observation}
\label{sec:impl-some-observ}
Figures~\ref{inaD3} and \ref{inaDsw3} presented the ring with high $\Sigma_{\rm d}$ at $r  \sim 1.5-2$ au.
We should note that the obtained ring can not be directly compared to rings observed in PPDs by ALMA \citep[e.g.,][]{ALMA-Partnership2015a, Andrews2018a} because the solid component already grows to $1$ km (the upper limit in our setting), which is not observed by millimeter/sub-millimeter wavelength.
In reality, however, the collisional fragmentation generates small dust grains that can be observed by ALMA.

The obtained rings are a possible source to supply crystalline materials to the outer disk.
Crystalline materials are detected at the outer surface layer of transitional/pre-transitional disks.
For instance, HD 100546 and HD 142527 are Herbig stars which have a circumstellar disk with a large gap or an inner cavity.
A large amount of crystalline forsterite is detected around $r\approx 13$-$20$ au of the disk in HD 100546 system \citep{Mulders2011}.
The crystalline water ice is also detected in the outer disk, $r\ge 146$ au around HD 142527 \citep{Min2016}.

An in-situ formation model cannot explain a large amount of these crystalline materials because the temperature in the outer region is too low to crystallize amorphous materials.
Therefore an additional heating mechanism or additional supply of crystalline materials is required.

The crystalline dust particles formed by the collision between planetesimals is one of the promising reservoirs of the detected crystalline materials \citep{Bouwman2003}.
Planetesimals can be heated enough to crystallize due to their accretion processes or disintegration of radioactive nucleus.
The collisional cascade of these planetesimals supplies a large amount of crystalline dust particles.
If a planetesimal ring is located in a gas starved region like an inner gap or an inner cavity, the crystalline dust particles are carried to the surface of the outer disk by the radiation pressure from the central star.

We suggest that MDW-dominated disks naturally serve suitable conditions to produce crystalline materials.
The MDW disperses the gas component from inside to outside with leaving a planetesimal ring with high $\Sigma_{\rm d}/\Sigma_{\rm g}$ in the inner region, as shown in Figures~\ref{inaD3} and \ref{inaDsw3}.
Such a dense ring has a high collision rate between planetesimals and supplies a large number of small particles even though there is no strong perturber.

\section{Conclusions}
\label{conclusions}

We investigated the evolution of dust grains in PPDs with the MDWs.
We calculated the evolution of the surface densities of the dust and gas components in PPDs under the 1+1 D (time + radial distance) approximation in various conditions of turbulent viscosity, the mass loss by the MDW, and the magnetic braking by the MDW.
We simultaneously solved a coagulation equation of solid particles under a single-size approximation with neglecting the effect of collisional fragmentation.

When the mass accretion by turbulent viscosity is efficient, most of the dust particles fall onto the host star before growing to the sufficiently large bodies even though the MDW is taken into account \added{(Fig.~\ref{aND3} and \ref{aD3})}.
The radial drift barrier is still a severe problem in PPDs governed by viscous accretion \added{in these cases}.

In contrast, however, if the mass accretion is moderately weak and the mass loss by the MDW is relatively important,
the radial dependence of the gas pressure is largely altered from the initial power-law profile by the MDW.
While all the dust grains drift inward to the central star, they are locally accumulated around the location of the minimum $\partial^2 \ln P/\partial (\ln r)^2(<0)$ because the dust flux is converging there.
When $\Sigma_{\rm d}/\Sigma_{\rm g} \gtrsim \eta$ in the drift limited state, St $\approx$ St$_{\rm eq} \gtrsim 1$, and then, the dust particles rapidly grow to planetesimal-sized objects by the positive feedback between the reduced radial drift and the accelerated collisional growth.

Once the dust grains reach the growth dominated phase, the mass loss of the solid component ceases from a PPD because the inward drift is halted at the ring of the solid concentration that is formed near the inner edge, $r=1.5$ au, of the dust disk.
The amount of the dust mass that is left in the disk is regulated by the time when St of the solid particles that constitute the ring exceeds unity.
This timing is determined by the combination of $\overline{\alpha_{r \phi}}, \overline{\alpha_{\phi z}}$, and $C_{\rm w}$.
The dependence on these parameters will be clarified in future studies.

This new growth mode of dust grains will be a promising mechanism for the formation of planetesimal.
Planetesimals formed by this process are distributed in a ring-like region with a clear inner edge.
The position of the planetesimal ring is supposed to determine the final outcome of the planetary system.
From observed orbital properties of exoplanets, we may infer the parameters of the MDW and turbulent viscosity of the PPDs. 

In addition, $\Sigma_{\rm d}$ of the ring region is much higher than the typical value of the MMSN model.
This implies the accelerated formation of protoplanets;
the rapid formation of gas-giant and inner rocky planets is further anticipated.
It would be important to study the planet formation in such local concentrations of planetesimals.

\acknowledgments
We are grateful to the anonymous referee for a careful reading of the manuscript and many valuable comments resulting in a substantial improvement of the original version of the paper.
We also thank Michiel Min, Douglas N. C. Lin, Sanemichi Z. Takahashi, and Eiichirou Kokubo for fruitful discussion.
This work was supported by Grants-in-Aid for Scientific Research from the MEXT of Japan, 17H01105, 17K05632, 17H01103, 18H05436, and 18H05438.
A part of simulations was carried out on PC cluster at Center for Computational Astrophysics, National Astronomical Observatory of Japan.


%\vspace{5mm}
%\facilities{HST(STIS), Swift(XRT and UVOT), AAVSO, CTIO:1.3m,


%\software{astropy \citep{2013A&A...558A..33A},  
%          Cloudy \citep{2013RMxAA..49..137F}, 
%          SExtractor \citep{1996A&AS..117..393B}
%          }



\appendix
\section{Single Size approximation}
\label{app:single_size_approximation}
The collisional evolution of the surface number density, $n_{\rm s}(m,r)$, of bodies with mass $m$ that rotate around a host star at $r$ is governed by \citep[e.g.,][]{Kobayashi2010b}
\begin{eqnarray}
 \frac{\partial m n_{\rm s}(m,r)}{\partial t} &=& \int_0^\infty dm_1
  \int_0^\infty dm_2 n_{\rm s}(m_1,r) n_{\rm s}(m_2,r)
  \times K(m_1,m_2) g(m,m_1,m_2) m_1 \nonumber \\
&& - m n_{\rm s} \int_0^\infty dm_2 n_{\rm s}(m_2,r) K(m_1,m_2)
-\frac{1}{r}[m r n_{\rm s}(m,r) v_{{\rm d},r}] 
- \int_0^\infty m \frac{n_{\rm s}(m,r)}{\sqrt{2\pi} h_{\rm d}} c_{\rm s} D_{\rm w} dm,
\label{eq:coag2}
\end{eqnarray}
where the first and second terms on the right-hand side are the contributions from collision, the third term denotes radial drift, and the fourth term indicates the mass loss dragged by MDWs.
Here, the collisional kernel between bodies with masses $m_1$ and $m_2$,
$K(m_1,m_2)$, is given by \citep{Okuzumi2012a}
\begin{equation}
 K(m_1,m_2) = \frac{\pi (a_1 + a_2)^2 \Delta v_{\rm p,1,2}}{\sqrt{2 \pi
  [h_{\rm d}(m_1)^2+h_{\rm d}(m_2)^2]}},
\end{equation}
where $g(m,m_1,m_2) dm$ is the number of bodies with masses ranging from $m$
to $m+dm$ produced from a single collision between $m_1$ and $m_2$ \citep[e.g.,][]{Kobayashi2010a},
$v_{{\rm d},r}$ is the radial drift velocity, $\Delta v_{\rm p,1,2}$
is the relative velocity of bodies with masses $m_1$ and $m_2$, and 
$h_{\rm d}$ is the scale height of solid grains.

We use $g(m,m_1,m_2) = \delta(m-m_1-m_2)$ for perfect sticking. 
Integrating eq.~(\ref{eq:coag2}) over $m$, we have
\begin{eqnarray}
 \frac{\partial \Sigma_{\rm d}}{\partial t} &=& 
-\frac{1}{r}\frac{\partial}{\partial r}[r \Sigma_{\rm d}
\langle v_{{\rm d},r} \rangle]
  - \frac{c_{\rm s}}{\sqrt{2 \pi}} \left\langle \frac{D_{\rm w}}{h_{\rm
  d}} \right\rangle \Sigma_{\rm d}, 
\label{eq:coag3}
\end{eqnarray}
where
\begin{eqnarray}
  \Sigma_{\rm d} &=& \int_0^\infty dm m n_{\rm s}(m,r), 
\\
\langle v_{{\rm d},r} \rangle
 &=& \frac{1}{\Sigma_{\rm d}} 
  \int_0^\infty dm v_{{\rm d},r} m n_{\rm s}(m,r), 
\end{eqnarray}
and
\begin{eqnarray}
\left\langle \frac{D_{\rm w}}{h_{\rm
  d}} \right\rangle &=& \frac{1}{\Sigma_{\rm d}} 
  \int_0^\infty dm \frac{D_{\rm w}}{h_{\rm d}} m n_{\rm s}(m,r).
\end{eqnarray}
Here $\langle \rangle$ indicates the mass-weighted average. 
In eq.~(\ref{eq:coag3}), 
the collisional terms are canceled out. 

Multiplying eq.~(\ref{eq:coag2}) by $m$, 
integrating it over $m$, and we then have 
\begin{eqnarray}
 \frac{\partial \langle m \rangle \Sigma_{\rm d}}{\partial t} = 
\langle K \rangle \Sigma_{\rm d}^2 
-\frac{1}{r}\frac{\partial}{\partial r}[r \langle m  v_{{\rm d},r}
\rangle \Sigma_{\rm d}]
- \frac{c_{\rm s}}{\sqrt{2 \pi}} \left\langle \frac{m D_{\rm w}}{h_{\rm
  d}} \right\rangle \Sigma_{\rm d},
\label{eq:coag_mass}
\end{eqnarray}
where 
\begin{eqnarray}
 \langle m \rangle = \frac{1}{\Sigma_{\rm d}} && \int_0^\infty dm m^2 n_{\rm s}(m,r), 
\\
 \langle m v_{{\rm d},r} \rangle = \frac{1}{\Sigma_{\rm d}} 
 && \int_0^\infty dm v_{{\rm d},r} m^2 n_{\rm s}(m,r), 
\\
 \langle K \rangle = \frac{1}{\Sigma_{\rm d}^2} \int_0^\infty dm_1 && \int_0^\infty
  dm_2 m_1 n_{\rm s}(m_1,r) 
 \times m_2 n_{\rm s} (m_2,r) K(m_1,m_2), 
\end{eqnarray}
and
\begin{eqnarray}
\left\langle \frac{m D_{\rm w}}{h_{\rm
  d}} \right\rangle = \frac{1}{\Sigma_{\rm d}} 
  \int_0^\infty dm \frac{m D_{\rm w}}{h_{\rm d}} m n_{\rm s}(m,r).
\end{eqnarray}

If we assume
\begin{eqnarray}
\langle m \rangle
 \langle v_{{\rm d},r} \rangle  = \langle m v_{{\rm d},r}
\rangle,\label{eq:assump1} \\
\langle m \rangle
\left\langle \frac{D_{\rm w}}{h_{\rm
  d}} \right\rangle =
\left\langle \frac{m D_{\rm w}}{h_{\rm
  d}} \right\rangle, \\
\langle K \rangle = \frac{2 \sqrt{\pi} a^2}{h_{\rm d}} \Delta v_{\rm pp},\label{eq:assump2}
\end{eqnarray}
and the averaged values are given by the values for $m_{\rm p}$, such as
$\langle m \rangle = m_{\rm p}$ and $ \langle v_{{\rm d},r} \rangle = v_{{\rm d},r}(m_{\rm p})$, eq.~(\ref{eq:coag3}) is then reduced to eq.~(\ref{eq:dust}).
Eqs.~(\ref{eq:coag3}) and (\ref{eq:coag_mass}) are then reduced to Eq.~(\ref{eq:growth}).
The averaged collisional kernel in eq.~(\ref{eq:assump2}) is determined by the collisional cross section and relative scale height between bodies with same mass $m_{\rm p}$, while $\Delta v_{\rm pp}$ is the relative velocity between bodies with ${\rm St}(m_{\rm p})$ and ${\rm St}(m_{\rm p})/2$ \citep{Sato2016a}.
Eqs.(\ref{eq:assump1}) and (\ref{eq:assump2}) are valid not only for the single size population but also for the growth until the onset of runaway growth \citep[e.g.,][]{Kobayashi2016a}.
Readers may find similar discussion in \citet{Sato2016a}.

\section{Derivation of condition for dust surface density enhancement}
\label{sec:deriv-cond-dust}
In this section, we derive the condition for the accumulation of solid particles by converging flow, which plays an important role in dust grains to reach the growth dominated phase.
We assume that the dust particles are in the drift limited phase of ${\rm St} \approx 0.1$ and that the surface density has a power-law profile of $\Sigma_{\rm d} \propto r^{-q_{\rm d}}$.

The dust mass flux is written as
\begin{eqnarray}
\label{eq:F}
 \dot{M}_{\rm d} = 2\pi r \Sigma_{\rm d} v_{{\rm d}, r},
\end{eqnarray}
where in the drift limited state the radial accretion speed of gas is generally much slower than the drift velocity of dust from gas, and therefore, we can safely neglect $v_{{\rm g},r}$ in eq.~(\ref{eq:vdr}).
We use $c_{\rm s}=c_{{\rm s},0}(r/r_0)^{-q_{T}/2}$, and $v_{\rm K}=v_{{\rm K},0}(r/r_0)^{-1/2}$, where $q_{T}$ is the power-law index of the temperature profile, as $T=T_{0}(r/r_{0})^{-q_{T}}$, and $r_0 = 1$ au.
We normalize $r$, $\Sigma_{\rm d}$, and $P$ by $r_{0}$, $\Sigma_{{\rm d}, 0}$, and $P_{0}$, respectively.
$\Sigma_{{\rm d},0}$ and $P_0$ are the dust surface density and the gas pressure at $r=r_0$.
Then, we obtain
\begin{eqnarray}
\label{eq:F2}
 \dot{M}_{\rm d} = 2\pi r_0 \Sigma_{{\rm d},0} v_{{\rm K},0} \left(\frac{c_{{\rm s},0}}{v_{{\rm K},0}}\right)^2\frac{{\rm St}}{1+{\rm St}^2} r^{\frac{3}{2}-q_{T}} \Sigma_{\rm d} \frac{r}{P}\frac{\partial P}{\partial r}, 
\end{eqnarray}
where $\dot{M}_{\rm d}$ is converging if $\mathrm{d}\dot{M}_{\rm d}/\mathrm{d}r < 0$ and dust particles are accumulated.

In the drift limited state, we can treat St as a constant value.
We introduce a normalization, $\dot{M}_{{\rm d},0} = 2\pi r_{0}\Sigma_{{\rm d}, 0}v_{{\rm K},0}(c_{{\rm s},0}/v_{{\rm K},0})^{2}{\rm St}/(1+{\rm St}^2)$, at $r=r_0$.
Then, the dimensionless divergence of the dust mass flux can be expressed as
\begin{eqnarray}
\label{eq:dF_dr}
\frac{\partial\dot{M}_{\rm d}}{\partial r} &=& 
\frac{\partial}{\partial r}\left[\left(r^{\frac{3}{2}-q_{T}} \Sigma_{\rm d}\right)\cdot\left(\frac{r}{P}\frac{\partial P}{\partial r}\right)\right] \nonumber \\
&=& r^{\frac{1}{2}-q_{T}}\Sigma_{\rm d}\left\{\frac{\partial^2 \ln P}{\partial (\ln r)^2} + \left[
\left(\frac{1}{2}-q_{T}\right)+
\frac{\partial \ln (r\Sigma_{\rm d})}{\partial \ln r}\right]\frac{\partial \ln P}{\partial \ln r}\right\}.
\end{eqnarray}
We use $q_{T}=1/2$ from MMSN model.
Therefore, if $\mathrm{d}\dot{M}_{\rm d}/\mathrm{d}r < 0$, then
\begin{eqnarray}
\label{eq:condition_exac}
\frac{\partial^2 \ln P}{\partial(\ln r)^2} <
- \frac{\partial \ln (r\Sigma_{\rm d})}{\partial \ln r}\frac{\partial \ln P}{\partial \ln r}.
\end{eqnarray}
This is the condition for dust grains to be piled up.

To examine the nature of this condition, we consider the situation that both the gas pressure and the dust surface density are expressed by power laws as $P=P_{0}(r/r_0)^{-q_{P}}$, $\Sigma_{\rm d}=\Sigma_{{\rm d},0}(r/r_0)^{-q_{\rm d}}$.
From eq.~(\ref{eq:condition_exac}), we obtain
\begin{eqnarray}
\label{eq:condition_power}
 q_{P}(1-q_{\rm d}) > 0.
\end{eqnarray}
In the inward (outward) drift cases with $q_{P}>0$ ($q_{P}<0$),  $q_{\rm d}<1$ ($q_{\rm d}>1$) is the condition for the increase of $\Sigma_{\rm d}$; in the opposite cases ($q_{\rm d}>1$ for $q_{P}<0$ or $q_{\rm d}<1$ for $q_{P}>0$), $\Sigma_{\rm d}$ decreases.
$\Sigma_{\rm d}$ is in the steady-state condition for $q_{\rm d}=1$.
This can be understood from the mass continuity equation (\ref{eq:dust}).
If we neglect the mass loss by the MDW, the variation of dust surface density is
\begin{eqnarray}
\label{eq:cont_eq_dust}
\frac{\partial \Sigma_{\rm d}}{\partial t} = - \frac{1}{r}\frac{\partial \dot{M}_{\rm d}}{\partial r}.
\end{eqnarray}
Here, $v_{{\rm d},r}$ (eq.~\ref{eq:vdr} with $v_{{\rm g},r}\ll v_{{\rm d},r}$) has a dependence of $v_{\rm d,r} \propto r^0$ because $\eta = -(h^2/r^2)\partial P/\partial r \propto r^{1/2}$ in the present setup.
Therefore, $\dot{M}_{\rm d} \propto r^{1-q_{\rm d}}$, which clearly shows that $q_{\rm d}=1$ is the steady-state condition.
In the inward drift condition of $q_{P}>0$, from eq.~(\ref{eq:cont_eq_dust}) we can obtain $\partial \Sigma_{\rm d}/\partial t \propto (1-q_{\rm d})/r^{1+q_{\rm d}}$.
Relative decrease (or increase) rate of $\Sigma_{\rm d}$ can be written as $\Gamma = (\partial \Sigma_{\rm d}/\partial t) / \Sigma_{\rm d} \propto r^{-1}$.
When the slope is steeper, $q_{\rm d}>1$, this relative decrease rate is faster for smaller $r$, and then, the slope gets shallower to $q_{\rm d}=1$.
Otherwise if $q_{\rm d}<1$, the relative increase rate is again faster for smaller $r$, which makes the slope steeper to $q_{\rm d}=1$.
The similar argument can be applied to the outward drift condition of $q_{P}<0$.
We can conclude that the slope of $\Sigma_{\rm d}$ tends to approach the steady-state value of $q_{\rm d}=1$ if the mass loss by the MDW can be neglected.

Finally, we consider more general cases in which the radial dependence of gas pressure is deviated from a simple power-law profile.
We assume $\Sigma_{\rm d}=\Sigma_{{\rm d},0}(r/r_0)^{-q_{\rm d}}$ again, and we can obtain eq.~(\ref{eq:pressure_condition}) from eq.~(\ref{eq:condition_exac}).




\bibliography{ref03_hk}

% \begin{thebibliography}{}
% \expandafter\ifx\csname natexlab\endcsname\relax\def\natexlab#1{#1}\fi
% \providecommand{\url}[1]{\href{#1}{#1}}
% \providecommand{\dodoi}[1]{doi:~\href{http://doi.org/#1}{\nolinkurl{#1}}}
% \providecommand{\doeprint}[1]{\href{http://ascl.net/#1}{\nolinkurl{http://ascl.net/#1}}}
% \providecommand{\doarXiv}[1]{\href{https://arxiv.org/abs/#1}{\nolinkurl{https://arxiv.org/abs/#1}}}

% \end{thebibliography}



\end{document}